\documentclass[10pt,a4paper]{article}
\usepackage{amsmath}
\usepackage{latexsym}
\usepackage{theorem}
\usepackage[all]{xy}
\newtheorem{teorema}{Theorem}[section]
\newtheorem{definicion}[teorema]{Definition}
\include{amslatex}
\newtheorem{proposicion}[teorema]{Proposition}

\newtheorem{corolario}[teorema]{Corollary}
\textheight
21 cm \textwidth 14 cm \oddsidemargin 10 mm \topmargin  0 mm
{\theorembodyfont{\rmfamily}
\newtheorem{comentario}[teorema]{Remark}}
{\theorembodyfont{\rmfamily}
\newtheorem{ejemplo}[teorema]{Example}}

\numberwithin{equation}{section}

\include{amsmath}
\begin{document}

\begin{title}
{\Large {\bf Fiber averaged dynamics associated with the Lorentz force equation}}
\end{title}
\maketitle
\author{

\begin{center}

Ricardo Gallego Torrom\'e\\
Department of Physics, Lancaster University,\\
Lancaster, LA1 4YB \& The Cockcroft Institute, UK\footnote{email: rigato39@gmail.com. Currently at the Departamento de Matem\'atica, Universidade Federal de S\~{a}o Carlos, Brazil.}
\end{center}}

\begin{abstract}
It is shown that the
 Lorentz force equation is equivalent to the auto-parallel condition
 $\,^L\nabla_{\dot{{x}}}\dot{{x}}=0$ of a linear connection
$^L\nabla$ defined on a convenient pull-back vector bundle. By using a geometric averaging
method, an associated {\it averaged Lorentz  connection} $\langle\,^L\nabla\rangle$ and
the corresponding auto-parallel equation are obtained. After this, it is shown that in the
ultra-relativistic limit and for narrow one-particle
probability distribution functions,
the auto-parallel curves of $\langle\,^L\nabla\rangle$ remain {\it nearby} close to the auto-parallel curves of
$^L\nabla$. Applications of this result in beam dynamics and plasma physics are briefly described.
\end{abstract}

\section{Introduction}
Given a four dimensional Lorentzian manifold $({\bf M},\eta)$,
the motion of a classical point charged particle under the influence of an external electromagnetic field and without taking into account radiation reaction is described by the {\it Lorentz force equation}.
This work explores two related topics:
\begin{enumerate}
 \item A geometric description of the Lorentz force equation and
 \item  An averaged version of the Lorentz force equation and its relation with the original Lorentz force equation.
\end{enumerate}
  Once a geometric version of the Lorentz force equation is available, we can apply the {\it averaging method} \cite{Ricardo}.
 It is proved that in the {\it ultra-relativistic limit} and under some natural assumptions on the averaging model, the solutions of the original Lorentz force equation can be approximated by the solutions of an {\it averaged Lorentz force} equation with high accuracy. This fact has interesting consequences for plasma modeling.

The geometrization of the Lorentz force equation is interesting from the point of view of the analysis of its symmetries and mathematical structure. It turns out that there are several connections whose geodesics equations are the Lorentz force equation \eqref{Lorentzequation} constrained by the speed normalization \eqref{metricconstraininLorentzforceequation} (see for instance \cite{Miron1} for one of such alternative connections). The structure of the Lorentz force equation \eqref{Lorentzequation} is described in terms of the Levi-Civita connection $^\eta\nabla$, the longitudinal and transverse tensors (see the tensor $L$ and $T$ discussed in {\it section} 3). The transverse tensor does not contribute to the geodesic equation but it does to the motion of a {\it charged gyroscope}, for example.

The {\it averaged Lorentz dynamics} has some advantages over the  {\it Lorentz dynamics}. The first and most notorious is that the differential equation
$\langle\,^L\nabla\rangle_{\dot{\tilde{x}}}{\dot{\tilde{x}}}=0$ is easier to work with than the original Lorentz force equation $\,^L\nabla_{\dot x} \dot x=0$. This allows us to perform easier analysis and computations with $\langle\,^L\nabla\rangle_{\dot{\tilde{x}}}{\dot{\tilde{x}}}=0$ than with $\,^L\nabla_{\dot x} \dot x=0$ (following the general philosophy of the averaging method in classical mechanics \cite{Arnold}). This is particularly useful when we apply the model to systems composed by a large number of point charged particles, composing a non-neutral plasma \cite{GT2,GT1}.

From a topological point of view, the averaged equation contains the same information than the original equation. This is because the connections $^L\nabla$ and $\langle\,^L\nabla\rangle$ are connected by a continuous homotopy, which is similar to the convex invariance found in Finsler symmetry \cite{Ricardo}. This property could have relevance for the study of the topological properties of confined plasmas, since could allow to study topological properties of plasmas by a simplified averaged model.

The structure of this work is the following.
In {\it section 2}, we explain how, given a second order differential equation, several connections are associated to the differential equation.

In {\it section 3}, a geometric interpretation for the Lorentz force equation is explained. In particular we
show how to extract  from the Lorentz force equation \eqref{Lorentzequation}, a linear connection defined on a convenient pull-back bundle. Such connection will be called the {\it Lorentz connection} $^L\nabla$. Then it is shown that equation $\eqref{Lorentzequation}$ corresponds to the auto-parallel condition of the Lorentz connection.

In {\it section 4}, we follow reference \cite{Ricardo, RF} and we define the {\it average of a family of automorphisms}, a generalization of the {\it fiber integration operation} of R. Thom and S. S. Chern \cite{BottTu}. Some examples of averaging in different geometric frameworks are explained. In {\it section 5}, given a manifold ${\bf M}$ and a subbundle $\hat{\bf N}$ of ${\bf TM}$, we apply the averaging method to an arbitrary linear connection on the pull-back bundle
$\pi^*|_{\hat{\bf N}}{\bf TM}$. The average of such connections are affine connections of the manifold {\bf M}. The averaging method is applied to the Lorentz connection,
obtaining the {\it averaged Lorentz connection} $\langle\,^L\nabla\rangle$.

In {\it section 6}, the solutions of the Lorentz force equation \eqref{Lorentzequation} and those of the auto-parallel
curves of the averaged connection are compared. The measure used to calculate the momentum moments in the averages are borrowed from  relativistic kinetic theory \cite{Ehlers}. Then it is proved that in the {\it ultra-relativistic limit},
for narrow $1$-particle probability distribution functions for
the same initial conditions, the
solutions of the auto-parallel equations $\,^L\nabla_{\dot x} \dot x=0$ and $\langle\,^L\nabla\rangle_{\dot{\tilde{x}}}{\dot{\tilde{x}}}=0$ remain {\it near} to each other, even for finite time evolution. We discuss the limits of validity of the result and provide two relevant examples of averaging for the Lorentz force equation.

In {\it section 7}, some applications in beam dynamics and plasma physics of the averaged Lorentz equation are briefly described, as well as other potential applications of the {\it averaged dynamics} and the extension to other equations of evolution.
In {\it section 8}, the hypothesis used in the proofs of the main results of this paper are discussed.

 \section{Connections and second order differential
equations}
Let $\pi:{\bf TM}\to {\bf M}$ be the canonical projection of the tangent bundle ${\bf TM}$ and $\Gamma {\bf TTM}$ the set of vector fields over ${\bf TM}$.
\begin{definicion}
A second order differential equation (or semi-spray) is a smooth vector field $G\in\Gamma\,{\bf TTM}$ such that $\pi_* G|_u=u,\,\forall \,u\in {\bf TM}$.
\label{definiciondesemiespray}
\end{definicion}
\begin{comentario}
If we require that the integral curves of the semi-spray $G$ are invariant under affine re-parameterization, it is necessary to exclude the origin $0\in \Gamma {\bf TM}$ from the domain of definition of $G$. This is because the invariance under affine re-parameterization implies that $G(x,y)$ must be homogeneous of degree one in $y$.
 The requirement that the integral curves are affine re-parameterization invariant implies that $G$ is a special type of second order differential equation known as spray. Therefore, a spray is not necessarily smooth at the origin.
 Moreover, since we will apply the geometric formalism to the Lorentz force differential equation, the associated geodesic spray is not defined over the whole ${\bf TM}\setminus\{0\}$, but only on the sub-bundle of timelike vectors fields ${\bf N}\to {\bf M}$ or on the future pointed unit hyperboloid sub-bundle ${\bf \Sigma}^+_x\to {\bf M}$, as we will define later. The particular geometric setting depends on the choice of parameterization for the curves describing the point particle. Therefore, it is convenient to develop the general geometric formalism for the case that $G$ is defined on an arbitrary open sub-bundle $\hat{\bf N}\hookrightarrow {\bf TTM}$, $G\in\,\Gamma T\hat{\bf N}$. In this case $\pi|_{\hat{\bf N}}:\hat{\bf N}\to {\bf M}$  is the restriction to $\hat{{\bf N}}$ of the canonical connection $\pi:{\bf TM}\to {\bf M}$.
 \end{comentario}

\subsection{{The non-linear connection associated to a second order differential equation}}

 Let us consider the differential map $\pi_*|_{\hat{\bf N}}:{\bf T}\hat{\bf N}\to {\bf TM}$. The vertical bundle (or vertical distribution) is the kernel $\mathcal{V}:=ker (\pi_*|_{\hat{\bf N}})$. At each point $u\in \hat{\bf N}$ one has $ker(\pi_*|{\hat{\bf N}}(u)):=\mathcal{V}_u$.
 \begin{definicion}
A non-linear connection on $\hat{\bf N}$ is a distribution $\mathcal{H}$ on ${\bf T}\hat{\bf N}$ such that
\begin{align*}
{\bf T}_u \hat{\bf N}=\mathcal{V}_u \oplus \mathcal{H}_u ,
\end{align*}
for each $u\in \,\hat{\bf N}$.
\end{definicion}
Given a semi-spray as in \ref{definiciondesemiespray},
there are connections partially characterized by the fact that by the projection $\pi|_{\hat{\bf N}}:\hat{\bf N}\to {\bf M}$, the auto-parallel curves coincide with the projection of the integral curves of $G$ \cite{Crampin, MironHrimiucShimadaSabau}. In order to introduce such connections, let us first consider the complete lift and vertical lift of tangent vectors \cite{Crampin}. The vertical lift $y^v$ of $y\in {\bf T}_x{\bf M}$ at the point $(x,\tilde{y})\in \,{\bf T}_x{\bf M}$ is the tangent vector at $s=0$ of the curve $s\mapsto (x,y+s\tilde{y})$. The complete lift of a vertical field $X\in\,\Gamma {\bf TM }$ is a vector field $X^c\in\,{\bf T N}$ whose flow in $\hat{\bf N}$ is $(s,(x,y))\mapsto (\phi_s(x),\phi_{s*}(y))$, where $ (s,x)\mapsto \phi_s(x)$ is the flow of $X$ and $\phi_{s*}$ is the flow of $X$ in ${\bf M}$. Given a semi-spray defined on the sub-bundle $\hat{\bf N}$ of ${\bf TM}$, the horizontal lift $y^h$ at the point $(x,\tilde{y})\in\,\hat{\bf N}$ is
\begin{align}
\iota_u:{\bf T}_x{\bf M}\to {\bf T}_u{\bf N},\quad y\mapsto \frac{1}{2}\big([X^v,G]|_{(x,\tilde{y})}+\,X^c|_{(x,\tilde{y})}\big),
\label{horizontalliftintrinsic}
\end{align}
where $X$ is any vector field on ${\bf M}$ such that $X|_x= y$. Then the set $\{\iota_u(y)|_{(x,{y})}\,|\,(x,y)\in \hat{\bf N}\}$.

 Let us consider the semi-spray $G\in \Gamma\,{\bf T\hat{N}}$. A semi-spray defined on $\hat{\bf N}$ (restriction from a semi-spray on ${\bf TM}$) can be expressed in a local frame associated to natural coordinates $({\bf TU},(x,y))$ on $\hat{\bf N}$ as
\begin{align}
G(x,y)=\,y^i\frac{\partial}{\partial x^i}-\,G^i(x,y)\frac{\partial}{\partial y^i},
\label{semisprayG}
\end{align}
such that the components $G^i(x,y)$ are smooth on  $\hat{\bf N}$. It is direct that he projection by $\pi|_{{\bf N}}:\hat{\bf N}\to {\bf M}$ of the integral curves of $G$ are solutions of the system of ordinary differential equations
\begin{align}
\frac{d y^i}{dt} -\, G^i(x,y)=0,\quad
\frac{d x^i}{dt}-\,y^i=0,\quad i=0,...,n-1
\label{firstorderdifferentialequations}
\end{align}
If the transformation in the local natural coordinates on the manifold $\hat{\bf
N}$ is of the form
\begin{displaymath}
\tilde{x}^i =\tilde{x}^i (x),\quad \tilde{y}^i=\frac{\partial\tilde{x}^i}{\partial x^j}y^j,
\end{displaymath}
 the corresponding time derivative transform as
 \begin{displaymath}
\frac{d\tilde{x}^i}{dt} =\frac{\partial \tilde{x}^i}{\partial x^j}\frac{dx^j}{dt},\quad
\frac{d\tilde{y}^i}{dt}=\frac{\partial^2 \tilde{x}^i}{\partial x^k \partial x^j}y^k\frac{
dx^j}{dt} + \frac{\partial \tilde{x}^i}{\partial x^j} \frac{dy^j}{dt}.
\end{displaymath}
This implies that in order to $G$ be a semi-spray, the functions $G^i(x,y)$ must transform as
\begin{align}
\tilde{G}^i(\tilde{x},\tilde{y})=\,\sum_{j,k,l,s} \big(\frac{\partial
\tilde{x}^l}{\partial x^j}\big)y^j\big(\frac{\partial
\tilde{x}^s}{\partial x^k}\big)y^k\frac{\partial^2
\tilde{x}^i}{\partial x^l \partial x^s} - \sum_j \big(\frac{\partial
\tilde{x}^i}{\partial{x}^j}\big){G}^j({x},{y}).
\label{transformationruleforGi}
\end{align}

Locally, the horizontal distribution $\mathcal{H}$ is generated by the frame
\begin{align}
\{\frac{\delta}{\delta x^0},..., \frac{\delta}{\delta
x^{n-1}}\},\quad\quad \frac{\delta}{\delta
x^k}:=\frac{\partial}{\partial x^k}-\frac{\partial{G^i}}{\partial
y^k} \frac{\partial}{\partial y^i},\quad i,j,k=0,...,n-1.
\label{horizontaldistribution}
\end{align}
Therefore, a  local frame for ${\bf T}\hat{\bf N}$ is given by
\begin{align}
\{ \frac{\delta}{\delta x^0},..., \frac{\delta}{\delta
x^{n-1}},\frac{\partial}{\partial y^0},...,\frac{\partial}{\partial y^{n-1}}\}.
\label{localframe}
\end{align}
and the dual local frame of $1$-forms is
\begin{align}
\{ dx^0,...,dx^{n-1},\,{\delta y^0},...,{\delta y^{n-1}}\}.
\label{localdualframe}
\end{align}
The duality relations are
\begin{align*}
dx^i(\frac{\delta}{\delta x^j})=\delta^i_j,\quad {\delta y^i}(\frac{\delta}{\delta x^j})=0,\quad
 dx^i(\frac{\partial}{\partial y^j})=0,\quad {\delta y^i}(\frac{\partial}{\partial y^j})=\delta^i_j,\quad i,j=0,...,n-1.
\end{align*}
The horizontal lift of the tangent vector $X =X^i\frac{\partial}{\partial x^i}|_x
\,\in {\bf T}_x{\bf M}$ to the tangent vector space ${\bf T}_u\hat{\bf N}$ is the tangent vector
$\iota_u(X)=X^i\frac{\delta}{\delta x^i}|_u$. Similarly, one can speak of horizontal lifts of vector fields.

\subsection{{Linear connections on the pull-back bundle $\pi^*|_{\hat{\bf N}}{\bf TM}$}}
Let us consider the cartesian product
$\hat{\bf N}\times{\bf TM} $ and the canonical projections
\begin{align*}
\pi _1:\hat{\bf N}\times{\bf TM} \to \hat{\bf N},\quad
(u,\xi)\mapsto u,\quad\quad
\pi _2 :\hat{\bf N}\times{\bf TM} \to {\bf TM},\quad
(u,\xi)\mapsto \xi.
\end{align*}
The pull-back bundle $\pi_1:\pi^*|_{\hat{\bf N}} {\bf TM}\to \hat{\bf N}$ of the bundle
$\pi:{\bf TM}\to {\bf M}$ by the projection $\pi:\hat{\bf N}\to {\bf M}$
 is the sub-manifold of the cartesian product
$\hat{\bf N}\times{\bf TM} $ such that the following equivalence
relation holds: for every $u\in {\bf N} $ and $(u,\xi) \in \pi^{-1}
_1 (u)$, $ (u,\xi)\in \pi^*|_{\hat{\bf N}}{\bf TM}$ {iff} $\pi \circ\pi
_2(u,\xi)=\hat{\pi}\circ \pi_1(u,\xi)$.
From the definition of the pull-back bundle $\pi|^*_{\hat{\bf N}}{\bf TM}$ it follows that the following diagram commutes,
\begin{displaymath}
\xymatrix{\pi^*|_{\hat{\bf N}}{\bf TM} \ar[d]_{\pi_1} \ar[r]^{\pi_2} &
{\bf TM} \ar[d]^{\pi}\\
\hat{\bf N} \ar[r]^{\pi} & {\bf M},}
\end{displaymath}
$\pi_1:\pi^*|_{\hat{\bf N}}{\bf TM}\to \hat{\bf N}$ is a real vector bundle,
with fibers over each $u=(x,y)\in \hat{\bf N}$ isomorphic to  ${\bf T}_x{\bf M}$.
The fiber dimension of $\pi^*|_{\hat{\bf N}}{\bf TM}$ is equal to $dim({\bf M})=n$, meanwhile the dimension of $\hat{\bf N}$ is $2n$.
A vector $Z\in {\bf T}_x{\bf M}$ can be pulled-back $\pi^*Z$ in a unique way by the following rules:
 \begin{itemize}
 \item $\pi_1(\pi^*Z)=(x,Z)\in \hat{\bf N}$,

 \item $\pi_2(\pi^*)=Z$.
 \end{itemize}
 The pull-back of a smooth function $f\in \mathcal{F}({\bf M})$ is a smooth function $\pi^*f\in \mathcal{F}(\pi^*|_{\hat{\bf N}}{\bf TM})$ such that $\pi^*f(u)=f(\pi_1(u))$ for every $u\in \hat{\bf N}$.
A local frame for the sections of $\pi^*|_{\hat{\bf N}}{\bf TM}$ is given by
  \begin{align*}
  \{\pi^*e_0,\pi^*e_1,....,\pi^*e_{n-1}\},
  \end{align*}
where $\{e_0,e_1,....,e_{n-1}\}$ is a local frame in ${\bf TM}$.
Analogous pull-back
bundles can be constructed from other tensor bundles
 over {\bf M}. Relevant examples are the pull-back bundle $\pi^*|_{\hat{\bf N}} {\bf T}^*{\bf M}\to \hat{\bf N}$ of the dual bundle ${\bf T}^*{\bf M}$ and more generally, the pull-back bundle $\pi^*|_{\hat{\bf N}}{\bf T}^{(p,q)}{\bf M}$ of tensor bundles ${\bf T}^{(p,q)}{\bf M}$, such that the following diagram commutes,
 \begin{displaymath}
\xymatrix{\pi^*|_{\hat{\bf N}}{\bf T}^{(p,q)}{\bf M} \ar[d]_{\pi_1} \ar[r]^{\pi_2} &
{\bf T}^{(p,q)}{\bf M} \ar[d]^{\pi}\\
\hat{\bf N} \ar[r]^{\pi} & {\bf M},}
\end{displaymath}
\begin{definicion}
A linear connection on the pull-back bundle ${\pi}_1:\pi^*|_{\hat{\bf N}}{\bf TM}\to \hat{\bf N}$ is a map
\begin{align*}
\nabla: \Gamma {\bf T}\hat{\bf N}\times  \Gamma\pi^*|_{\hat{\bf N}}{\bf TM}\to \Gamma\pi^*|_{\hat{\bf N}}{\bf TM}
\end{align*}
such that
\begin{itemize}
\item For every $X\in \Gamma {\bf T}\hat{\bf N}$, $S_1, S_2\in \Gamma\pi^*|_{\hat{\bf N}}{\bf TM}$ and $f\in\mathcal{F}({\bf M})$, it holds that
\begin{align}
\nabla_X (\pi^*f S_1+S_2)=\,(X(\pi^* f))S_1+\,f\nabla_X S_1+ \,\nabla_X S_2.
\label{leibnitzrule}
\end{align}

\item For every $X_1, X_2\in \Gamma {\bf T}\hat{\bf N}$, $S\in \Gamma\pi^*|_{\hat{\bf N}}{\bf TM}$ and $\lambda\in\mathcal{F}(\hat{\bf N})$, it holds that
\begin{align}
\nabla_{f X_1+X_2}S=\,f\nabla_{X_1}S+\,\nabla_{X_2}S.
\label{Xlinearity}
\end{align}
\end{itemize}
\label{connectioninpullback}
\end{definicion}
\subsection{Linear connections on $\pi^*|_{\hat{\bf N}}{\bf TM}$ associated to a semi-spray $G$.}
Let us define the following functions,
 \begin{align}
^G\Gamma^i \,_{jk}(x,y):=\frac{1}{2}\, \frac{\partial^2
G^i(x,y)}{\partial y^j \partial y^k}, \quad i,j,k=0,...,n-1.
\label{Ggammacoefficients}
\end{align}
The behavior under natural coordinate transformations of the functions $^G\Gamma^i \,_{jk}$  is analogous to the transformation rules for the connection coefficients of a linear connection on ${\bf M}$. However, they do not live on ${\bf M}$ but on $\hat{\bf N}$. To overcome this difficulty, one considers linear connections on the pull-back bundle $\pi^*|_{\hat{\bf N}}{\bf TM}$ as follows,
\begin{proposicion}
 Given a semi-spray $G(x,y)$, there is a unique linear connection $^G\nabla$ in $\pi^*|_{\hat{\bf N}}{\bf TM}$ such that in the locally frames $\{\frac{\delta}{\delta x^0},...,\frac{\delta}{\delta x^{n-1}},\frac{\partial}{\partial x^0},...,\frac{\partial}{\partial x^{n-1}}\}$ and $\{\pi^*e_i, \,i=0,...,n-1\,\}$ the following covariant derivatives,
\begin{align}
^G\nabla_{\frac{\delta}{\delta x^j}} \,\pi^* Z:=
\,^{G}\Gamma(x,y)^i\,_{jk}\,Z^k\,\pi^*e_i,\quad\quad ^G\nabla_{V}
\pi^* Z:=0,
\label{structuralequationsforG}
\end{align}
for every $\quad V\in \mathcal{V},\,Z\in \,\Gamma{\bf TM}$.
\label{propositiononnablaG}
\end{proposicion}
{\bf Proof}.
It is clear that the homomorphism defined by the conditions \eqref{structuralequationsforG} satisfies the axioms of a connection as in definition \ref{connectioninpullback},
 \begin{itemize}
\item $^G\nabla_{\lambda X} Z$ is linear in $\lambda$ and $X$, as in condition \eqref{Xlinearity}.

\item For every $f\in \,\mathcal{F}({\bf M})$, it holds that
\begin{align}
^G\nabla _{{ X}} \pi^* (fS):= \big({X}(\pi^*f\big))\,\pi^* S+\,\big(\pi^*f\big)\,\,^G\nabla _{{ X}} \pi^* S \quad \forall {X}\in {\bf T}_u\hat{\bf
N}, \,f\in\,\mathcal{F}({\bf M}),\,S\in\,\Gamma \,{\bf TM}.
\label{leibnitzruleforgnabla}
\end{align}
\end{itemize}
\label{connectionassociatedtoG}
The Leibnitz rule \eqref{leibnitzruleforgnabla} follows from the definition of pull-back of functions. Other linearity properties follow easily as well.  Uniqueness follows from the linearity and Leibnitz rule.
 \hfill$\Box$

Recall that the torsion tensor of an affine connection $D$ of ${\bf M}$ is a tensor defined by the expression
 \begin{align}
Tor_{D}(X,Y)=\,D _{X} Y-\, D _{Y} X -[X,Y],\quad X,Y\in \,\Gamma {\bf TM}.
\label{torsionofaaffineconnection}
\end{align}
The {\it generalized torsion} of a linear connection ${\nabla}$ on $\pi^*|_{\hat{\bf N}}{\bf TM}$ is the linear  homomorphism
\begin{align*}
Tor_{{\nabla}}:\Gamma \,{\bf TM}\times\, \Gamma \,{\bf TM}\to \Gamma\,\pi^*|_{\hat{\bf N}}{\bf TM},
\end{align*}
defined by the formula
\begin{align}
Tor_{{\nabla}}(X,Y)=\,\nabla _{\iota _u (X)} \pi^*|_{\hat{\bf N}}   Y -\, |_u  \nabla _{\iota _u (Y)} \pi^*|_{\hat{\bf N}}  X -\pi^*|_{\hat{\bf N}}[X,Y],\quad  X,Y\in \,\Gamma {\bf TM},
\label{generalizedtorsion}
\end{align}
where $u\in \hat{\bf N}$ is the evaluating point. Thus, for instance, it is direct from the definition of pull-back that $\pi^*|_{\hat{\bf N}}([X,Y]|_x)|_u$ is an element of the fiber $\pi^{-1}_1(x)$. Varying $x\in \,{\bf M}$, $\pi^*([X,Y])$ defines a {\it section} of $\pi^*|_{\hat{\bf N}}{\bf TM}$. Similarly, to the tangent vector $X|_x\in\,{\bf T}_x{\bf TM}$ and to the section $Y\in \,\Gamma {\bf TM}$, we associate the covariant derivative $,\nabla _{\iota _u (X)} \pi^*|_{\hat{\bf N}}   Y $ and similar for other terms.
In a similar way, the {\it mixed torsion} is defined as the linear homomorphism
\begin{align}
K_\nabla (V,Z):\Gamma \,\mathcal{V}\times\, \Gamma \,{\bf TM}\to \Gamma\,\pi^*|_{\hat{\bf N}}{\bf TM},\quad
K_{\nabla}(V,Z):=\nabla_{V}
\pi^*|_{\hat{\bf N}} Z,\quad V\in \mathcal{V},\,Z\in \,\Gamma{\bf TM}.
\end{align}
A connection $\nabla$ on $\pi^*|_{\hat{\bf N}}{\bf TM}$ is called {\it torsion-free} iff the generalized torsion and the mixed torsion are zero.
Then the following {\it proposition} is direct,
\begin{proposicion}
The connection $^G\nabla$ as in {\it proposition} \ref{propositiononnablaG} is torsion-free.
\label{propositiontorisionfree}
\end{proposicion}
The space of linear and {\it torsion-free} connections on $\pi^*|_{\hat{\bf N}}{\bf TM}$ is denoted by $\nabla_{\hat{\bf N}}$.

The covariant derivative $^G\nabla$ can be extended to the pull-back $\pi^*{\bf T}^{(p,q)}{\bf M}$  of
tensor bundles ${\bf T}^{(p,q)}{\bf M}$ over ${\bf M}$ by assuming that the product rule holds:
\begin{align*}
^G\nabla_X (f\,T_1\otimes\,T_2)=\,f\,^G\nabla_X T_1\otimes \,T_2+\,T_1\,\otimes \,^G\nabla_X T_2+\,(X( f))\,T_1\,
\end{align*}
for any pair of tensors $T_1$ and $T_2$
and for any smooth function $f$ on $\hat{{\bf N}}$. Finally, the covariant derivative along $X$ of a function $f\in\, \mathcal{F}(\hat{{\bf N}})$ is defined as the function on $\hat{{\bf N}}$ given by the directional derivative
\begin{align*}
^G\nabla_X f:=\,X(f).
\end{align*}

\section{{Geometric formulation of the Lorentz force equation}}
Let us consider a Lorentzian metric $\eta$  on the four dimensional manifold ${\bf M}$ with signature $(1,-1,-1,-1)$.
\begin{definicion}
A Lorentzian Randers space is a triplet $({\bf M},\eta, {\bf F})$, where $({\bf M},\eta)$ is a Lorentzian manifold and ${\bf F}$ is a closed $2$-form on ${\bf M}$.
\label{Lorentzianranderspace}
\end{definicion}

By Poincar\'e's lemma, for any point $x\in {\bf M}$, there is a locally smooth $1$-form $A$ such that in a neighborhood of $x$, the relation $dA=F$ holds. Therefore, a Lorentzian Randers space can be equivalently described by a triplet $({\bf M}, \eta, [A])$, with $[A]$ being the equivalence class of all locally smooth $1$-forms whose exterior derivative, on the domains where they are smooth, is the $2$-form $F$.

Given a Lorentzian Randers space $({\bf M},\eta, {\bf F})$,  a natural Lagrangian function can be defined locally
\begin{align}
\mathcal{L}=\,\sqrt{\eta(y,y)}+\,A(y),
\label{lagrangianofarandersspace}
\end{align}
where in $x\in {\bf U}$, $(x,y)\in \,{\bf TU}$, $y\in\,{\bf T}_x{\bf U}$ is timelike ($\eta(y,y)>0$) and $A(y)$ is the action of the $1$-form $A$ on the vector $y\in \,{\bf T}_x{\bf M}$. Note that although this Lagrangian is locally defined, the associated Euler-Lagrange equations are globally defined, since only depend on the $2$-form $F$ and the metric $\eta$.
The restriction to timelike vectors such that $\eta(y,y)>0$ makes natural to describe  geometry of the Lorentz force equation \eqref{Lorentzequation} can be described  in the framework of the theory explained in {\it sections 2}. We first consider the convex bundle of timelike vectors,
\begin{align}
\pi|_{\bf N}:{\bf N}\to {\bf M},\quad {\bf N}:\bigsqcup_{x\in{\bf M}} {\bf N}_x,\quad {\bf N}_x:=\,\{y\in {\bf T}_x{\bf M},\,\,s.t.\,\,\eta(y,y)>0\},
\label{timelikecone}\end{align}
where $\pi|_{\bf N}$ is the restriction of the canonical projection $\pi$ to ${\bf N}$. The fiber over each point $x\in {\bf M}$ is the open cone of timelike vectors at $x$ and has dimension $n$.  ${\bf N}$ is clearly a subbundle of the tangent bundle ${\bf TM}$.

When parameterized by a parameter such that $\eta(y,y)$ is constant, the timelike geodesics of the Lagrangian \eqref{lagrangianofarandersspace} coincide with the solutions of the Lorentz force equation \eqref{Lorentzequation}. Furthermore, not that although $F=dA$ holds only locally in general, a gauge transformation $A\to A+d\phi$ leaves the equations of motion $\eqref{Lorentzequation}$ invariant.

If the curves are parameterized by the proper time associated with $\eta$, the geometric theory of the Lorentz force equation must be formulated on the unit tangent hyperboloid bundle $\pi|_{\bf \Sigma}:{\bf \Sigma}\to {\bf M}$, where $\pi|_{\bf \Sigma}$ is the restriction of $\pi|_{\bf N}$ to
\begin{align}
\pi|_{\bf \Sigma}:{\bf \Sigma}\to {\bf M},\quad {\bf \Sigma}:\bigsqcup_{x\in{\bf M}} {\bf \Sigma}_x,\quad {\bf \Sigma}_x:=\,\{y\in {\bf T}_x{\bf M},\,\,s.t.\,\,\eta(y,y)=1\}.
\label{definitionofsigma}
\end{align}
We formulate first the geometric theory of Lorentz force equation for bundles and objects over  ${\bf N}$ and then restrict to bundles and objects over  ${\bf \Sigma}$. This is motivated by physical applications, specially in {\it section 6}, where the additional constraint of being all the world-lines future pointed will be imposed.

There are several reasons to consider the non-affine formulation of the Lorentz force equation, instead of directly the proper time formulation:
\begin{itemize}
\item Since the description of the motion of a charged particle is performed on arbitrary coordinate systems using arbitrary time parameters, it is convenient to consider the arbitrary parameterized equation \eqref{Lorentzequation}.
\item The Lorentz force equation \eqref{Lorentzequation} is the condition for the critical points for the action functional with Lagrangian \eqref{lagrangianofarandersspace}. This admits a geometric interpretation in terms of {\it conic Finsler spaces} \cite{JavaloyesSanchez}.

\item It allows in a natural way to define the covariant derivative of vector fields out from the unit hyperboloid ${\bf \Sigma}$. This is of significant when comparing geodesics.

\item It admits a natural reduction to vector fields living on ${\bf \Sigma}$.
\end{itemize}
These reasons make natural to consider the re-parameterization invariant formulation of the Lorentz force equation instead of the affine parameterized invariant.
\subsection{{The Lorentz force as a non-linear geodesic equation associated to a spray}}
{\bf Notation}. It will be convenient for the calculations that will follow to introduce some further notation. Let
let $({\bf M}, \eta,F)$ be a Lorentzian Randers space.
 For an arbitrary $1$-form $\omega$ we denote by ${\omega}^{\sharp}:=\eta^{-1}(\omega,\cdot)$
the vector obtained by duality, using the Lorentzian metric $\eta$. Given a general $2$-form $\beta$, the $(1,1)$ tensor $\beta^\sharp$ is defined by the relation
\begin{align*}
 (\beta^\sharp)(\theta,Y)=\,\beta (\theta^\sharp,Y),\quad \theta \in \,\Lambda^1 {\bf M},\,X \in \Gamma {\bf TM}.
 \end{align*}
  Similarly, given
a vector field $X$ over {\bf M}, one can define the dual $1$-form ${X}^*:=\eta(X,\cdot)$;
$X \cdot\omega$ is the inner product of the vector $X$ with the $k$-form $\omega$, giving a $(k-1)$-form. On ${\bf TN}$, there is a canonical tangent structure,
\begin{align}
J:{\bf TN}\to {\bf TN},\quad J=\sum^{n-1}_{k=0}\,\frac{\partial}{\partial y^k}\otimes\,dx^k.
\label{tangentstructure}
\end{align}
This homomorphism is extended to tensors of type $(1,1)$ on ${\bf N}$.
 It is useful to introduce the  homomorphism,
\begin{align}
\tilde{J}_1:\Gamma {\bf TN}\to \Gamma {\bf T}^{(1,1)N},\quad \tilde{X}\mapsto \frac{1}{2}\,\sum^{n-1}_{0}dx^i\otimes \,[\frac{\partial}{\partial y^i},\tilde{X}].
\label{homomorphismoJ0}
\end{align}
Thus, each derivation respect to a vertical direction of a spray coefficients is associated with the action of the operator $\tilde{J}_1$ on the spray itself.

The Lorentz force equation is the affine re-parameterization invariant second order differential equation
\begin{align}
^\eta\nabla_{\dot{x}}\,\dot{x} +\,\sqrt{\eta(\dot{x},\dot{x})}\,(\dot{x}\cdot F)^\sharp=0
\label{Lorentzequation}
\end{align}
for $t\in {I}=[a,b]$.
$^{\eta}\nabla$ are the coefficients of the
Levi-Civita connection $^{\eta}\nabla$ of $\eta$ and $F=dA$ is the exterior derivative of the $1$-form $A$.

The system of second order differential equations \eqref{Lorentzequation} determines a spray vector field on ${\bf N}$. Indeed, it is useful to perform some calculations to consider the  non-parameterized by the proper-time world-lines by the proper parameter of $\eta$. The normalization condition
 \begin{align}
\quad\eta(\dot{x},\dot{x})=1,
\label{metricconstraininLorentzforceequation}
\end{align}
where $\dot{x}$ is the tangent vector to the curve, is imposed only at the end of the derivative operations.

The Lorentz spray vector field on {\bf N}  associated with the Lorentz force equation  is defined to be the vector field $^LG\in \,\Gamma T{\bf N}$
given by
\begin{align}
^LG=\, ^{\eta}G+\,\sqrt{\eta(y,y)}\,J((y\cdot F)^\sharp),
\label{spraylorentz}
\end{align}
where $^\eta G$ is the Riemannian spray of $\eta$. Note that $^LG-\,^\eta G$ is vertical.
The spray \eqref{spraylorentz} was proposed first by R. Miron \cite{Miron1}.
\begin{proposicion}
The non-linear connection associated to a Lorentzian Randers space $({\bf M},\eta,F)$ is obtained from the spray \eqref{spraylorentz} and the Levi-Civita connection of $\eta$.
\end{proposicion}
{\bf Proof}. It is enough to check that, when written in local coordinates, the expression \eqref{Lgammacoeffients} determines the non-zero connection coefficients \eqref{Ggammacoefficients} defined by the spray \eqref{spraylorentz}.
This is because the components of the tensor
\begin{align}
 \tilde{J}_1\big(\,^LG-\, ^{\eta}G\big)=\,\frac{1}{2}\,\Big(\frac{1}{\sqrt{\eta(y,y)}}y^*\otimes J((y\cdot F)^\sharp)+\,J(F^\sharp) \sqrt{\eta(y,y)}\Big)
\label{nonlinearconnectioncoefficientsofLG}
\end{align}
 determine the components of the non-linear connection  $^LN$ in terms of coefficients of the non-linear connection associated to the Levi-Civita connection of $^\eta G$ and in terms of the two $2$-form $F$.
 \hfill$\Box$

It is immediate, by application of Euler's theorem of homogeneous functions, that the integral curves of $^LG$ correspond to the solutions of the Lorentz force equation \cite{Miron1}.

\subsection{{The Lorentz force equation as auto-parallel condition of connections on the pull-back $\pi^*|_{\bf N}{\bf TM}$}}
Let us consider the pull-back bundle $\pi_1:\pi^*|_{\bf N}{\bf TM}\to {\bf N}$, such that the following diagram commutes:
\begin{displaymath}
\xymatrix{\pi^*|_{\bf N}{\bf TM} \ar[d]_{\pi_1} \ar[r]^{\pi_2} &
{\bf TM} \ar[d]^{\pi}\\
{\bf N} \ar[r]^{\pi} & {\bf M}.}
\end{displaymath}
In this subsection, a geometric version of the Lorentz force equation is formulated using a linear connection on $\pi^*|_{\bf N}{\bf TM}$. The vertical bundle is $\mathcal{V}:=\,ker\,\pi_*|_{\bf N}$.  One can introduce the following homomorphism,
\begin{align}
\tilde{J}_2:\Gamma {\bf TN}\to \Gamma {\bf T}^{(1,2)}{\bf N},\quad \tilde{X}\mapsto \pi^*|_{\bf N}\Big(\sum^{n-1}_{j,k=0} dx^j\otimes dx^k\otimes [\frac{\partial}{\partial y^j}[\frac{\partial}{\partial y^k},\tilde{X}]]\Big).
\label{homomorphismoJ1}
\end{align}

The spray coefficients  $^L\Gamma^i\,_{jk}$  are obtained from the components of the tensor $(1,2)$ tensor
\begin{align*}
\frac{1}{2}\tilde{J}_2\big(\,^LG -\,^\eta G \big) & = \,\Big(-J((y\cdot F)^\sharp)\frac{1}{2({\eta(y,y)})^{3/2}}y^*\otimes y^*+\,\frac{1}{2}\frac{1}{\sqrt{\eta(y,y)}}\,J(F^\sharp)\otimes y^*  \\
& +  \,\frac{1}{2}\frac{1}{\sqrt{\eta(y,y)}}\,J((y\cdot F)^\sharp)\otimes \eta
+\, \frac{1}{2}\frac{1}{\sqrt{\eta(y,y)}}\,y^*\otimes J(F^\sharp)\Big),
\end{align*}
The structure of the connection $^L\nabla$ is the following: $^{\eta}\nabla$ of the Levi-Civita connection of the Minkowski metric $\eta$ (seen as a non-linear connection associated to the spray $^\eta G$. There are two relevant tensorial terms in the connection, given by the expressions
\begin{align*}
 L=\,\frac{1}{2{\sqrt{\eta(y,y)}}}\, Sym(J(F^\sharp) (x)\otimes y^*),\,\quad
 T=\,\frac{1}{2{\sqrt{\eta(y,y)}}} \,J((y\cdot F)^\sharp)(x)\otimes\,
(\eta-\frac{1}{\eta(y,y)}y^*\otimes y^*)\big),
\end{align*}
where
 \begin{align*}
 Sym(J(F^\sharp) (x)\otimes y^*) =\,J(F^\sharp) \otimes y^*+\,y^*\otimes J(F^\sharp) .
 \end{align*}
 For each tangent vector
 $y\in {\bf T}_x{\bf M}$ with $\eta(y,y)>0$, we have the relation
\begin{align}
^LG-\,^{\eta}G=
\frac{1}{2{\sqrt{\eta(y,y)}}}\,Sym(J(F^\sharp) \otimes y^*) +
J((y\cdot F)^\sharp)\otimes\,
(\eta-\frac{1}{\eta(y,y)}y^*\otimes y^*).
\label{Lgammacoeffients}
\end{align}
A fundamental property of the transverse tensor $T$ is that
\begin{align}
y\cdot T=0,\quad \forall\, y\in\,{\bf T}_x{\bf M}.
\label{fundamentalpropertyofT}
\end{align}
Analogously, a fundamental property for the tensor $L$ is that
\begin{align}
y\cdot (y\cdot L)=\,\frac{1}{{\sqrt{\eta(y,y)}}}\,J((y\cdot F)^\sharp)(x),\quad \forall\, y\in\,{\bf T}_x{\bf M}.
\label{fundamentalpropertyofT}
\end{align}
The associated pull-back connection of  \eqref{Lgammacoeffients} on the unit hyperboloid bundle  ${\bf \Sigma}$ induced by the natural embedding $e:{\bf \Sigma}\to {\bf N}$ is
\begin{align}
^LG-\,^{\eta}G=\,\Big(
\frac{1}{2}\big(Sym(J(F^\sharp) \otimes y^*) +
J((y\cdot F)^\sharp)\otimes\,
(\eta-y^*\otimes y^*)\big)\Big)
\label{lorentzconnectioncoefficientsonsigma}
\end{align}
\begin{proposicion}
The following properties hold:
\begin{enumerate}
\item The connection $^L\nabla$ are invariant under gauge
transformations $A\to A+d\lambda$ of the $1$-form $A(x)$.

\item For each $x\in \,{\bf M}$, there is a local coordinate system centered at $x$ which coincides with a normal coordinate system associated
with the Levi-Civita connection $^{\eta}\nabla$ of the Lorentzian metric $\eta$ iff $F(x)=0$.
\end{enumerate}
\end{proposicion}
{\bf Proof}.
\begin{enumerate}
\item
The first property is direct from the fact that all the geometric
objects appearing in the definition of $^L\nabla$ are gauge invariant.

\item If we assume that there is a normal coordinate system centered at $x$ for $^L\nabla$ and
that this coordinate system coincides with the normal coordinate
system associated with $^{\eta}\nabla$.
Then by the {\it transversality condition} \eqref{fundamentalpropertyofT} one has the relation
\begin{displaymath}
0=y\cdot (y\cdot L),\quad \forall \,y\in{\bf N}_x.
\end{displaymath}
This is equivalent to
\begin{align*}
y\cdot{F}^\sharp=0,\,\quad\forall y\in {\bf N}_x.
\end{align*}
Therefore, in such coordinate system, one has that ${F}^\sharp=0$.
\hfill$\Box$
\end{enumerate}
\begin{proposicion}
There is an unique linear connection $^L\nabla$ on the
pull-back bundle $\pi^*|_{\bf N}{\bf TM}\rightarrow {\bf N}$ determined by the following structural equations:
\begin{enumerate}
\item It is a {\it torsion-free} connection,
\begin{enumerate}
\item The following relation holds,
\begin{align}
Tor_{^L\nabla}(X,Y)=\,^L\nabla_{\tilde{X}} \pi^*|_{\bf N} Y -\,^L\nabla_{\tilde{Y}}\pi^*|_{\bf N} X
-\pi^*|_{\bf N}[X,Y]=0,
\label{torsioncondition1}
\end{align}
where $X,Y\in \Gamma{\bf TM}$ are timelike and $\tilde{X},\tilde{Y}\in \Gamma{\bf
TN}$ are horizontal lifts to $\Gamma {\bf TN}$.

\item The mixed torsion $K_{^L\nabla}$ is zero,
\begin{align}
K(V,e_k)=\,^L\nabla _{V} \pi^*|_{\bf N}e_k=0,\quad V\in \mathcal{V},\, k=0,...,n-1.
\label{torsioncondition2}
\end{align}
\end{enumerate}
\item The covariant derivative along horizontal directions is
given by the formula
\begin{align}
^L\nabla _{\frac{\delta}{\delta x^j}} \pi^*|_{\bf N} e_k=\, ^L{\Gamma}^i\, _{jk}(x,y)\,\pi^*|_{\bf N}
e_i,\quad (i,j,k=0,...,n-1).
\label{derivationalonghorizontal}
\end{align}
\end{enumerate}
\label{proposiciononLnablaonpistartm}
\end{proposicion}
{\bf Proof.}
A general covariant derivative can be expressed in terms of the connection $1$-forms:
\begin{displaymath}
\omega^i\,_j(x,y):=\,^L\Gamma^i\,_{jk}(x,y)\,dx^k +\, ^L\Upsilon^i\,_{jk}(x,y)\,{\delta y^k},
\end{displaymath}
From the fact that the covariant derivative of sections on $\pi^*|_{\bf N}{\bf TM}$ along vertical directions is zero, one obtains
\begin{displaymath}
^L\Upsilon^i\,_{jk}(x,y){\delta y^k}=0\, \Rightarrow\, ^L\Upsilon^i\,_{jk}(x,y)=0
\end{displaymath}
at each point $(x,y)\in {\bf N}$.
The derivation $\pi^*|_{\bf N}{\bf TM}$ along horizontal directions
 is defined by  equation \eqref{derivationalonghorizontal}. Since the coefficients given by
formula \eqref{Lgammacoeffients} are symmetric in the lower indices, this rule is consistent with the torsion-free condition.
\hfill$\Box$

For a covariant derivative operator, the Leibnitz rule must be satisfied. Therefore, covariant derivative of $\pi^*|_{{\bf N}} (fS)$ must be
\begin{align}
^L\nabla _{{ X}} \pi^*|_{{\bf N}} (fS)=\big({X}\cdot \pi^*|_{{\bf N}}f\big)\,\pi^*|_{{\bf N}} S+\,\big(\pi^*|_{{\bf N}}f\big)\,\,^L\nabla _{{ X}} \pi^*|_{{\bf N}} S \quad \forall {X}\in {\bf T}_u{\bf
N}, \,f\in\,\mathcal{F}({\bf M}),\,S\in\,\Gamma \,{\bf TM}.
\label{lderivativeofafunction}
\end{align}
The covariant derivative $^L\nabla$ satisfies \eqref{lderivativeofafunction}. This is proved  by direct computation; for a vertical vector field $V\in \mathcal{V}$, one has that
\begin{align*}
0=\,^L\nabla _{{ V}} \pi^*|_{{\bf N}} (fS) &=\, \big({V}\cdot \pi^*|_{{\bf N}}f\big)\,\pi^*|_{{\bf N}} S+\,\big(\pi^*|_{{\bf N}}f\big)\,\,^L\nabla _{{ V}} \pi^*|_{{\bf N}} S\\
& =\,0\, \pi^*|_{{\bf N}} S +\pi^*|_{{\bf N}}f\cdot 0=0.
\end{align*}
Similarly, for any vector field $H\in \mathcal{H}$, one has that
\begin{align*}
^L\nabla _{{ H}} \pi^*|_{{\bf N}} (fS) &=\,(d(\pi^*|_{{\bf N}} f))(H)\, \pi^*|_{{\bf N}} S+\,\pi^*|_{{\bf N}}f( \,^L\nabla _{{ H}}\pi^*|_{{\bf N}} S)\\
& = \big({H}\cdot \pi^*|_{{\bf N}}f\big)\,\pi^*|_{{\bf N}} S+\,\big(\pi^*|_{{\bf N}}f\big)\,\,^L\nabla _{{ H}} \pi^*|_{{\bf N}} S,
\end{align*}
where the first line is because the Leibnitz's rule that a linear connection hold and the second equality is because the relation $d\pi^*|_{{\bf N}} f\cdot H={H}( \pi^*|_{{\bf N}}f$).
\begin{proposicion}
The auto-parallel curves of the linear Lorentz
connection $^L\nabla$  on the pull-back bundle $\pi^*|_{\bf N}{\bf TM}$ are in one to one correspondence with the solutions
 of the Lorentz force equation.
 \label{relationbetweenlorentzequationandLnabla}
\end{proposicion}
{\bf Proof}. Let us assume that in local coordinate system, the connection coefficients of the Lorentz connection $^L\nabla$
 has the connection coefficients $^L\Gamma^i\,_{jk}(x,y)$ on the bundle $\pi^*|_{\bf N}{\bf TM}$ are
\begin{align*}
\,^L\nabla_{\iota\big(\dot{x}\big)}\,&\pi^* \dot{x}\, =\,\pi^*\Big(\,^\eta\nabla_{\iota\big(\dot{x}\big)}\,\pi^* \dot{x}\Big)+\,\dot{x}\cdot\big(\dot{x}\cdot L\big)+\dot{x}\cdot\big(\dot{x}\cdot T\big)\\
& =\,\pi^*\Big(\,^\eta\nabla_{\iota\big(\dot{x}\big)}\,\pi^* \dot{x}\Big)+\,\dot{x}\cdot\big(\dot{x}\cdot L\big)\\
& =\,\pi^*\Big(\,^\eta\nabla_{\iota\big(\dot{x}\big)}\,\pi^* \dot{x}\Big)+\,\frac{1}{2{\sqrt{\eta(\dot{x},\dot{x})}}}\,\dot{x}\cdot\Big(\dot{x}\cdot\big( Sym(J(F^\sharp) (x)\otimes \dot{x}^*\big)\Big)\\
& =\,\pi^*\Big(\,^\eta\nabla_{\iota\big(\dot{x}\big)}\,\pi^* \dot{x}\Big)+\,\sqrt{\eta(\dot{x},\dot{x})}\big(\dot{x}\cdot F\big)^\sharp,
\end{align*}
This is  the Lorentz force equation \eqref{Lorentzequation}.
\hfill$\Box$

\section{{The averaging operator associated with bundle morphisms}}
 In this {\it section} we recall the general formalism of averaging geometric objects defined in pull-back vector bundles.   We will follow the formulation of the averaging operation introduced in \cite{Ricardo, RF}, in particular for the averaging operation of {\it families bundles automorphisms} of the pull-back bundle $\pi^*|_{\hat{\bf N}}{\bf T}{\bf M}$, where $\pi|_{\hat{\bf N}}:{\hat{\bf N}}\to {\bf M}$ is a sub-bundle of ${\pi}:{\bf TM}\to {\bf M}$.
 Let $\pi _1 ,\pi _2 $ be the canonical
projections of the pull-back bundle $\pi^*|_{\hat{\bf N}} {\bf T}{\bf
M}\rightarrow {\bf N}$, $\pi:{\bf T}{\bf M}\to {\bf M}$. $S_x$ is a
generic element of the fiber $\pi^{-1}(x)\subset \,{\bf T}{\bf M}$ and $S_u$ is the
evaluation of the section $S\in\,\Gamma \big(\pi^*|_{\hat{\bf N}} {\bf T}{\bf M}\big)$ at the point
$u\in\hat{\bf N}$ and therefore $S_u\in \,\pi^{-1}_1(u)$. With this notation, the following diagram commutes,
\begin{displaymath}
\xymatrix{S_u \ar[d]_{\pi_1} \ar[r]^{\pi_2} &
S_x \ar[d]^{\pi}\\
u \ar[r]^{\hat{\pi}} & x.}
\end{displaymath}
If $S_x\in {\bf T}_x{\bf M}$, it can be expressed in a local frame as
$S_x=\,S^i(x)\frac{\partial}{\partial z^i}\big|_x$, then the pull-back over the point $u=(x,y)\in \,\pi^{-1}(x)$ is the fiber element
$S_u=\,S^i(x)\pi^*|_{\bf \hat{N}}\frac{\partial}{\partial z^i}\big|_u$. This tensor is covariant under local coordinates changes on the fiber induced by local change of coordinates on {\bf M} \cite{BCS}.
\subsection{{Average of bundle automorphisms of $\pi^*|_{\hat{\bf {N}}}{\bf T}{\bf M}$}}
{\bf Admissible measures}. Let us denote the vector space of polynomials of degree $p$ or less on the variable components $y\in\,\hat{\bf N}_x$ by $Pol_x(y)$.
\begin{definicion}
A {\it measure of degree p} on each fiber $\hat{\pi}^{-1}(x)={\bf N}_x$ of ${\bf TM}$ is a smooth family of $p$-form  distributions $\{\omega_x\in
\mathcal{D}^*({\bf N}_x),\,x\in\,{\bf M}\}$ with the following compactness property: the smooth function on $\hat{\bf N}$,
\begin{align*}
Mess:Pol_x(y)\to {\bf R}^+,\quad P_p(x,y)\mapsto \int_{\hat{\bf N}_x}\,P_p(x,y)\,\omega_x(y)
\end{align*}
 is finite for any polynomial function  $P_p(x,y)\in\,Pol_x(y)$.
\label{definitionofmeasurep}
\end{definicion}
The averaging operation requires the following elements to be defined,
\begin{enumerate}
\item A sub-bundle $\pi|_{\hat{\bf N}}:\hat{\bf N}\to {\bf TM}$,

\item A {\it measure of degree p} on each fiber ${\pi}^{-1}(x)$ for each $x\in \hat{\bf M}$.

\item A fiber embedding $e_x:\hat{{\bf N}}_x\to \pi^{-1}(x)$, for each $x\in\,\bf M$.
\end{enumerate}
\begin{ejemplo}{\bf Examples of admissible measures.}
\begin{itemize}
\item A family of volume forms $\{\omega_x\, x\in\,{\bf M}\}$ with compact support on each $\hat{{\bf N}}_x$

\item Measure relevant for relativistic models are tempered distributions  with support on $\hat{\bf N}_x$ \cite{Hoermander}.
\end{itemize}
 \end{ejemplo}
 Given an admissible measure, the volume function $vol(\hat{\bf N}_x)$ is defined by
\begin{align*}
vol:{\bf M}\to {\bf R},\quad x\mapsto vol(\hat{\bf N}_x):=\,\int_{\hat{\bf N}_x}1\,\omega_x(y).
\end{align*}
For an admissible measure, the volume function is finite at each $x\in {\bf M}$.
{\bf Averaging operation}.
From the properties of the measure $\omega$, it follows that $vol(\hat{\bf N}_x)$ is smooth on ${\bf M}$. We will also assume that it is  bounded on {\bf M}.

For each tensor $S_z \in {\bf
T}^{(p,q)} _z{\bf M}$ with $\, v\in
\pi^{-1}(z)$, $z\in {\bf U}\subset {\bf M}$, the following isomorphisms are defined:
\begin{align*}
\pi _2 | _v :\pi^{-1}_2(v) \to \pi^{-1}(x),\quad S_w\mapsto S _z,\quad
\pi^*|_v :\pi^{-1}(x)\to \pi^{-1}_2(v)  ,\quad S _z\mapsto  \pi^*
_v S_z .
\end{align*}
Consider a family of automorphism
${\bf A}:=\big\{A_w:\pi ^*_w {\bf TM} \to \pi^*_w {\bf TM},\, w\in\,\hat{\bf N}_x\big\}$.
The action of the family of automorphism ${\bf A}$ is the sub-set$ \{A_u\pi^*_u S_x,u\in\pi^{-1}(x)\}$ of the bundle  $\pi^*|_{\hat{\bf N}}{\bf T}_x{\bf M}\to \hat{\bf N}_x$. Along such submanifold one can perform fiber integration (see for instance \cite{BottTu} for a general fiber integration on vector bundles). Then one can define the following fiber valued integrals,
\begin{align}
\Big(\int _{\hat{\bf N}_x}
 \pi
_2 |_u  A_u  \pi^*_u \Big)\cdot\,S_x := \int _{\hat{\bf N}_x}\big(
 \pi
_2 (  A_u  \pi^* _u \,S_x)\,\omega_x(u) \big),\,\,\forall S_x
\in {\bf T}_x {\bf M}.
\label{fiberintegration}
\end{align}
The right hand side integral is a fiber integration on $\hat{\bf N}_x$. Thus, the chain of compositions is
\begin{align}
x\mapsto \,S_x\to \{\pi^*_u S_x\}\to \{A_u(\pi^*_u S_x)\}\to \{\pi_2(A_u(\pi^*_u S_x))\}\to \int_{\hat{\bf N}_x} \pi_2(A_u(\pi^*_u S_x))\omega_x.
\end{align}
\begin{definicion}
 The average operator of the family of automorphism ${\bf A}$ is the family of automorphisms $\langle {\bf A}\rangle:=\{\langle A\rangle_x ,\,x\in \,{\bf M}\}$ where each automorphism $\langle A\rangle_x  : {\bf T}_x  {\bf M} \to  {\bf T}_x{\bf M}$ is defined as
 \begin{align*}
\langle A\rangle_x  : {\bf T}_x & {\bf M} \to  {\bf T}_x{\bf M}\\
& S_x \mapsto \frac{1}{vol(\hat{\bf N}_x)}\Big(\int _{\hat{\bf N}_x}
 \pi
_2 |_u  A_u  \pi^* _u \Big)\cdot\,S_x ,\,\quad u\in {\pi^{-1}_2(x)},\,\forall \, S_x
\in {\bf T}_x {\bf M}.
\end{align*}
\label{averagingoperation}
\end{definicion}
We will denote average operators by symbols between angles. Note that the
averaged operation is not canonically defined, since it depends on the measure $\omega_x$.
\begin{comentario}In order to define this average operation, the elements $\pi_2 A_u  \pi^* _u \cdot\,S_x$ must be possible to be added. Thus, it is required an addition operator for the elements $\pi_2 A_u  \pi^* _u \cdot\,S_x$.
\end{comentario}

{\bf Average operation on sections}.
The averaging operation can also be defined for a family of operators acting on sections of the pull-back bundle $\pi^*|_{\hat{\bf N}}{\bf TM}$.
Let $\pi _1 ,\pi _2 $, $\pi^*|_{\hat{\bf N}} {\bf TM}$ and ${\bf TM}$ be as before. Let
us consider the sections $S\in \Gamma{\bf TM}$,
$\pi^*|_{\hat{\bf N}}S\in \Gamma \pi^*|_{\hat{\bf N}} {\bf TM}$ and
the following isomorphisms,
\begin{align*}
\pi _2 :\pi^*|_{\hat{\bf N}}  {\bf TM}\to
{\bf T}^{(p,q)} {\bf M},\quad S_v\mapsto S _z,
\end{align*}
and
\begin{align*}
\pi^*|_{\hat{\bf N}} :{\bf TM}\to
\pi^*|_{\hat{\bf N}}{\bf TM} ,\quad S _z\mapsto \pi^*
_v S_z .
\end{align*}
Thus, the average of a family of operators acting on sections is defined pointwise:
\begin{definicion}
The average of an automorphism
\begin{align*}
A:\Gamma\,\pi^*|_{\hat{\bf N}} {\bf TM} \to \Gamma\,\pi^*|_{\hat{\bf N}} {\bf TM}
\end{align*}
is the is the automorphism
\begin{align*}
\langle A\rangle  : \Gamma\,{\bf TM} \to  \Gamma\,{\bf TM}\\
& S_x \mapsto \frac{1}{vol(\hat{\bf N}_x)}\Big(\int _{\hat{\bf N}_x}
 \pi
_2 |_u  A_u  \pi^* _u \Big)\cdot\,S_x ,\,\quad u\in {\pi^{-1}(x)},\, S_x
\in {\bf T}^{(p,q)}_x {\bf M}.
\end{align*}
\label{averagingoperation2}
\end{definicion}
{\bf Two fundamental properties of the averaging}
Let us consider two local frames $\{\pi^*|_{\hat{\bf N}}e_i(u),\,i=0,...,n-1\}$ and $\{\pi^*|_{\hat{\bf N}}\tilde{e}_i(u),\,i=0,...,n-1\}$ for sections of $\pi^*|_{\hat{\bf N}}{\bf TM}$. Under coordinate transformations, the components of the fiber element
 $S(v)=\tilde{S}^i(v)\pi^*|_{\hat{\bf N}}\tilde{e}_i(v)$ transform as
\begin{align}
\tilde{S}^i(v)=\frac{\partial \tilde{x}^i}{\partial {x}^k}S^k(v),
\label{tSi}
\end{align}
where $v\in \pi^{-1}(x),\,x\in{\bf M}$.
\begin{proposicion}
The averaging of $A$ is a geometric operation, independent of the local coordinate system on {\bf M}.
 \end{proposicion}
  {\bf Proof}. With the transformation law \eqref{tSi} for the components of $ S $ under local coordinate changes on {\bf M}, the averaging operation is independent of the coordinates:

\begin{align*}
\langle A\rangle (S(x)) & =\frac{1}{vol(\hat{\bf N}_x)}\Big(\int _{\hat{\bf N}_x}
 \pi
_2 |_u  A_u  \pi^* _u \Big)\cdot\,S(x)\\
& = \frac{1}{vol(\hat{\bf N}_x)}\Big(\int _{\hat{\bf N}_x}
 \pi
_2 |_u  A_u  \pi^* _u \Big)\cdot\,{S}^i(x) e_i(x)\\
& = {S}^i(x)\frac{1}{vol(\hat{\bf N}_x)}\Big(\int _{\hat{\bf N}_x}
 \pi
_2 |_u  A_u  \pi^* _u \Big)\cdot\,\frac{\partial \tilde{x}^k}{\partial {x}^i}\tilde{e}_k(x)\\
& ={S}^i(x)\frac{\partial \tilde{x}^k}{\partial {x}^i}\frac{1}{vol(\hat{\bf N}_x)}\Big(\int _{\hat{\bf N}_x}
 \pi
_2 |_u  A_u  \pi^* _u \Big)\cdot\,\tilde{e}_k(x)\\
& = \tilde{S}^k(x)\frac{1}{vol(\hat{\bf N}_x)}\Big(\int _{\hat{\bf N}_x}
 \pi
_2 |_u  A_u  \pi^* _u \Big)\cdot\,\tilde{e}_k(x)\\
& = \frac{1}{vol(\hat{\bf N}_x)}\Big(\int _{\hat{\bf N}_x}
 \pi
_2 |_u  A_u  \pi^* _u \Big)\cdot\,\tilde{S}^k(x)\tilde{e}_k(x).
\end{align*}
Comparing the first line with the last line in the above calculation, we obtain the result.\hfill$\Box$
\begin{proposicion}
The averaged of the family the family of automorphisms
\begin{align*}
\{\widetilde{Id}:\Gamma\,\pi^*|_{\hat{\bf N}} {\bf TM} \to \Gamma\,\pi^*|_{\hat{\bf N}} {\bf TM},\,\tilde{S}_u\mapsto \tilde{S}_u\}
\end{align*}
is the identity operator
\begin{align*}
Id:\Gamma\,\pi{\bf TM}\to \Gamma\,\pi{\bf TM} ,\quad S_x\to S_x.
\end{align*}
\label{identityoperator}
\end{proposicion}
{\bf Proof}. It can be shown by a direct calculation:

\begin{align*}
\langle \widetilde{Id} \rangle (S(x)) & =\frac{1}{vol(\hat{\bf N}_x)}\Big(\int _{\hat{\bf N}_x}
 \pi
_2 |_u  \widetilde{Id}_u  \pi^* _u \Big)\cdot\,S(x)\\
& =\frac{1}{vol(\hat{\bf N}_x)}\Big(\int _{\hat{\bf N}_x}
 \pi
_2 |_u \pi^* _u \Big)\cdot\,S(x)\\
& =\frac{1}{vol(\hat{\bf N}_x)}\Big(\int _{\hat{\bf N}_x}
 \pi
_2 |_{(x,y)} \pi^* _{(x,y)} S(x)\omega_x(y) \Big)\\
& \frac{1}{vol(\hat{\bf N}_x)}\Big(\int _{\hat{\bf N}_x} S(x)\omega_x(y) \Big)\\
& \frac{1}{vol(\hat{\bf N}_x)}\Big(\int _{\hat{\bf N}_x}\omega_x(y) \Big) S(x)=S(x).
\end{align*}
\hfill$\Box$
\subsection{Examples of averaging operations of relevance for the Lorentz force dynamics}
one has the following examples of admissible measures and averaging operations,
\begin{itemize}
\item {\it Lorentzian Structure}. The geometric data is a Lorentzian
metric $\eta$ on {\bf M}. The family of disjoint sub-manifolds ${\bf N}_x\subset {\bf T}_x{\bf M}$ are chosen to be the timelike vectors at each point $x\in\,{\bf M}$.
The measure on each fiber ${\bf N}_x$ is given by the $n$-form
\begin{align}
\omega_x(y):=f(x,y)\sqrt{\det \eta}\,dy^1\wedge\cdot\cdot\cdot\wedge dy^{n-1},
\label{vol}
\end{align}
where $f_x(y)=f(x,y)$ is positive and with compact support on ${\bf
N}_x$ and $\det\eta$ is the determinant of the matrix components of $\eta$.

Given a Lorentzian structure $({\bf M},\eta)$, there are several natural choices of sub-bundles $\hat{\bf N}$ that are natural to take the average operation. For instance, one can consider $\hat{{\bf N}}$ to be the collection of null cones over {\bf M},
\begin{displaymath}
{{\bf LC}}:=\bigsqcup_{x\in {\bf M}}\{{{\bf LC}}_x \subset{\bf T}_x{\bf M}\setminus\{0\}\},\quad {{\bf
LC}}_x:=\{y\in {\bf T}_x{\bf M}\setminus\{0\}\, | \,\eta(y,y)=0\}.
\end{displaymath}
$\pi:{\bf LC}\to {\bf M}$ is the light-cone bundle over {\bf M} and ${\bf LC}_x$ is
the light-cone over $x$. On  the other hand, $e:{\bf LC}\hookrightarrow {\bf TM}$ is a subbundle of ${\bf TM}$.
With these elements, one can perform averages of geometric objects in the unit hyperboloid or on the light-cone respectively. The admissible measure used is the same as in covariant kinetic theory \cite{Ehlers}.

\item
Let us consider a Lorentzian manifold $({\bf M},\eta)$, where the spacetime manifold ${\bf M}$ is a $4$-dimensional and the metric $\eta$ has signature $(+,-,-,-)$. A time orientation in $M$ is a smooth timelike vector field $U$. If this is possible,  $M$ is called time orientable. Then one can define the bundle
\begin{align}
{\bf \Sigma}^+:=\bigsqcup_{x\in {\bf M}}\{{\bf \Sigma}^+_x \subset{\bf T}_x{\bf M}\,\},\quad{\bf N}^+_x:=\{y\in {\bf T}_x{\bf M}\, | \,\eta(y,y)=1,\,\eta(y,U)>0\}.
\label{futurepointedtimelikecone}
\end{align}
The measure on ${\bf \Sigma}^+_x$ is determined by
the $(n-1)$-form
$\omega_x(y):=f(x,y)\,dvol_y(x)$ (equation \eqref{vol} before), with the constraint that the support of $f(x,y)$ on ${\bf
N}^+_x$ is compact. This example will be of physical relevance in {\it section 6}, where we will apply the average to the dynamics of on-shell point particles.
\end{itemize}

\section{The averaged Lorentz connection}
In this {\it section} we formulate first the averaging operation for connections on the vector bundle $\pi^*|_{\hat{\bf N}}{\bf TM}\to \hat{\bf N}$, where $\pi|_{\hat{\bf N}}:{\hat{\bf N}}\to {\bf M}$ is a $0$-codimension subbundle of $\hat{\bf TM}$. Then we apply the averaging to the connection $^L\nabla$ on  bundle $\pi^*|_{\bf N}{\bf TM}$, where ${\bf N}$ is timelike cone bundle. Finally, we investigate the induced connection on the unit hyperboloid of future point points $\pi_{{\bf \Sigma}^+}:{\bf \Sigma}^+\to {\bf M}$.
\subsection{Averaged of a linear connection on $\pi^*|_{\hat{\bf N}}{\bf TM}$}
Let us consider the pull-back bundle $\pi|_{\hat{\bf N}}:\pi^*|_{\hat{\bf N}}{\bf TM}\to \hat{\bf N}$.
The volume form that we use in performing integrals along the fibers is \eqref{vol}, that we write in the form
\begin{align}
\omega_x(y):=f(x,y)\, dvol(x,y).
\label{volumeformforomega}
\end{align}
The forms $\{\omega_x(y),\,x\in{\bf M}\}$ define a form $\omega_x(y)=\omega(x,y)\in \Gamma \,\Lambda^{n}\hat{\bf N}$ such that
the required integrals in the definition of the Lorentz averaged connection are finite.
\begin{proposicion}
Let the bundle $\pi_{\hat{\bf N}}:\hat{\bf N}\to {\bf M}$ be endowed with a non-linear connection
 and let us consider a linear connection $\nabla$ defined on
the pull-back bundle $\pi^*|_{\hat{\bf N}}{\bf TM}\to \hat{\bf N}$. Then there
is defined an affine connection of ${\bf M}$ determined by the
covariant derivative of $Y$ in the direction $X$ as
\begin{align}
\langle\nabla\rangle_X  Y:= \langle \pi _2|_u
{\nabla}_{\iota(X)} \pi^* _v Y\,\rangle ,\,\, \forall \,\,v\in {\bf U}_u
\label{averagedconnectionformula}
\end{align}
for each $ X\in {\bf T}_x {\bf M}$ and $Y\in\,\Gamma\, {\bf TM}$ with ${\bf U}_u$ being an open neighborhood of $u$.
\label{averagedconnection}
\end{proposicion}
{\bf Proof}.
We check that the properties for a linear covariant
derivative hold for $\langle \nabla\rangle$.
\begin{enumerate}
\item The operator $\langle\nabla\rangle_X$ is a linear application acting on
vector sections of $ {\bf M}$:
\begin{equation}
\langle\nabla\rangle_{X}(Y_1 +Y_2 )=\langle\nabla\rangle_{X} Y_1
+\langle\nabla\rangle_{X} Y_2,\quad
\langle\nabla\rangle_{X} \lambda Y= \lambda\langle\nabla\rangle_{X} Y,
\quad \forall \,\,  Y_1 , Y_2, Y\in {\bf \Gamma M}, \lambda\in
 {\bf R}, \quad X\in {\bf T}_x {\bf M}.
\end{equation}
For the first equation, the proof consists in the following
calculation:
\begin{align*}
\langle\nabla\rangle_{X }(Y_1 +Y_2 ) &=\langle\pi _2|_u {\nabla}_{\iota(X)} \pi^* _v 
(Y_1 +Y_2 )\rangle\\
& =\langle\pi _2 |_u  {\nabla}_{\iota(X)} \pi^* _v Y_1
\rangle+\langle\pi _2 |_u   {\nabla}_{\iota(X)}  \pi^* _v Y_2 \rangle\\
& =\langle\nabla\rangle_{X} Y_1 +\langle\nabla\rangle_{X} Y_2 .
\end{align*}
For the second condition:
\begin{align*}
\langle\nabla\rangle_{X } ({\lambda}Y)=\langle\pi _2 |_u {\nabla}_{\iota(
X)}
\pi^*_v ({\lambda}Y)\rangle ={\lambda}\langle\pi _2 |_u  {\nabla}_{\iota(X)} 
\pi^*_v \rangle =\lambda\langle\nabla\rangle_{X } Y.
\end{align*}

\item $\langle\nabla\rangle_{X } Y$ is a ${\bf \mathcal{F}}$-linear
respect $X$:
\begin{align*}
\langle\nabla\rangle_{X_1 +X_2 } Y=\langle\nabla\rangle_{X_1 }
Y+\langle\nabla\rangle_{X_2 } Y,\quad\quad \langle\nabla\rangle_{fX} (Y)=f(x)\langle\nabla\rangle_{X } Y,
\end{align*}
\begin{equation}
\forall \, Y\in M,v\in {\pi}^{-1}(z),\quad X,X_1,X_2 \in {\bf T}_x {\bf M},\, 
f\in{\bf \mathcal{F}}({\bf M}).
\end{equation}
To prove the first equation it is enough the following
calculation:
\begin{align*}
\langle\nabla\rangle_{X_1 +X_2 }Y & =\langle \pi _2 |_u ({\nabla}_{\iota
(X_1 +X_2 )})\pi^*_v Y\rangle \\
& =\langle\pi _2 |_u {\nabla}_{\iota( X_1) }\pi^*_v Y \rangle +\langle\pi _2
|_u {\nabla}_{\iota(X_2) }\pi^*_v Y \rangle\\
& =(\langle\nabla\rangle_{X_1 } Y)+(\langle\nabla\rangle_{X_2 } Y).
\end{align*}

For the second condition the proof is similar.

\item The Leibnitz rule holds:
\begin{align}
\langle\nabla\rangle_{X }
(fY)=df(X)Y+f\langle\nabla\rangle_{X } Y,\quad \forall \,\,  Y\in {\bf M},\, f\in \mathcal{F}{\bf 
M},\quad X \in {\bf T}_x {\bf M},
\label{Leibnitz}
\end{align}
 where $df(X)$ is the action of the $1$-form $df\in \,\Gamma \,{\Lambda}^1 _x {\bf M}$ on $X\in 
{\bf T}_x {\bf M}$. In order to prove \eqref{Leibnitz} we use the following
property:
\begin{displaymath}
{\pi}^*_v (f Y)={\pi}^*_v f{\pi}^*_v Y,\quad \forall \,\, Y\in {\bf
TM},\, f\in \mathcal{F}({\bf M}).
\end{displaymath}
Then one obtains the following expressions,
\begin{align*}
\langle{\nabla}\rangle_{X} (f Y) & =\langle \pi _2 |_u {\nabla}_{\iota(
X)}\pi^*_v (f Y)\rangle =\langle \pi _2 |_u {\nabla}_{\iota(X)}
{\pi}^*_v (f)\pi^*_v Y\rangle\\
& =\langle \pi _2 |_u (\nabla _{\iota(X)}( \pi^* _v f))\pi^*_v (Y)\rangle+\langle\pi _2 
|_u (\pi^* _u f){\nabla}_{\iota(X)}\pi^*_v (Y)\rangle\\
& =\langle\pi _2 |_u  (\,{\iota(X)}({\pi}^*_v f)) \pi^*_v (Y)\rangle + f_x \langle\pi _2 |_u 
{\nabla}_{\iota(X)} \pi^*_v (Y)\rangle \\
& =\langle (X_x f)\pi _2 |_u \pi^*_u (Y)\rangle +f_x \langle\pi _2 |_v
{\nabla}_{\iota(X)}\pi^*_v (Y)\rangle .
\end{align*}
For the first term we perform the following simplification,
\begin{align*}
\langle(X f)\pi _2 |_u \pi^*_u (Y)\rangle  =(X f)\langle \pi _2 |_u  \pi^*_u 
(Y)\rangle =(X f)(\langle  \pi _2 |_u \pi^*_u \rangle )Y=(X f)Y .
\end{align*}
Finally we obtain that
\begin{align*}
\langle{\nabla}\rangle_{X} (f Y) =\tilde{\nabla} _{X_x}(f)Y+f
\langle{\nabla}\rangle_{X}Y =df(X)Y+f\langle\nabla\rangle_{X }Y.
\end{align*}
\hfill$\Box$
\end{enumerate}
 From the definitions of generalized torsion tensor $Tor_{\nabla}$ \eqref{generalizedtorsion} and of torsion of an affine connection \eqref{torsionofaaffineconnection}, we have that
\begin{proposicion}
Let {\bf M} be an $n$-dimensional manifold and $\nabla$ a linear
connection on the bundle $\pi^*|_{\hat{\bf N}} {\bf TM}\to {\bf M}$
with generalized torsion $Tor_{\nabla}$. Then
\begin{align}
Tor_{\langle{\nabla}\rangle}(X,Y)=\,\langle Tor_{\nabla}\rangle(X,Y).
\label{formulafortheaveragedtorsio}
\end{align}
\end{proposicion}
{\bf Proof}. We follow reference \cite{Ricardo}.
Given for arbitrary vector fields $X,Y
\in \Gamma {\bf  TM}$, the torsion tensor $T_{\langle\nabla\rangle}$ on $X,Y$ is given by
\begin{align*}
Tor_{\langle\nabla\rangle}(X,Y) & =\,\langle\pi _2 |_u \nabla_{\iota(X)} \pi^* _w \rangle Y\, -\langle\pi 
_2 |_u  \nabla_{^h Y} \pi^* _w \rangle X -[X,Y]\\
&
=\,\langle \pi _2 |_u \nabla _{\iota(X)}  \pi^* _u \rangle Y -\,\langle\pi _2 |_u 
\nabla _{\iota( Y)} \pi^* _u \rangle  X -\langle \pi _2 |_u  \pi^* _u [X,Y]\rangle\\
& =\langle \pi _2 |_u  \big(\nabla _{\iota(X)}\pi^* Y -\nabla _{\iota(Y)}\pi^*X 
-\pi^*[X,Y] \big) \rangle\\
& =\,\langle Tor_{\nabla}\rangle(X,Y).
\end{align*}
\hfill$\Box$

\begin{corolario}
Let {\bf M} be an $n$-dimensional manifold and
$\nabla$ a linear connection on $\pi^*|_{\hat{\bf N}}{\bf TM}$ with unique non-zero
connection coefficients $\Gamma^i\,_{jk}(x,y)$ to be defined by the expression
\begin{align}
\nabla_{^h e_j} \pi^* e_k =\Gamma^i\, _{jk} \pi^* e_i,\quad
i,j,k=0,1,...,n-1.
\label{nablaofgamma}
\end{align}
Then the averaged connection $\langle\nabla\rangle$ is determined by the covariant derivatives
\begin{align}
\langle \nabla_{(e_j)} \rangle e_k =\langle\Gamma\rangle\,^i\, _{jk}  e_i,\quad
i,j,k=0,1,...,n-1.
\label{nablaofgammaaveraged}
\end{align}
with the associated  connection coefficients
\begin{align}
\langle\Gamma\rangle^i\,_{jk}:=\langle{\Gamma}^i\, _{jk}\rangle(x)= \frac{1}{vol(\hat{\bf N}_x)} \int_{e(\hat{\bf
{N}}_x)}\,\, \Gamma^ i\,_{jk}(x,y)\,\omega_x(y),\quad
i,j,k=0,1,...,n-1.
\label{gammapromediado}
\end{align}
\label{corolariosobregammapromedio}
\end{corolario}
{\bf Proof.} Let $\{e_i\}$, $\{\pi^* e_i\}$ and $\{\,^he_i\}$ be
local frames for the vector bundles ${\bf TM}$,
$\pi|_{{\hat{\bf N}}}^*{\bf TM}$ and the horizontal bundle $\mathcal{H}$ such
that the connection coefficients $\Gamma ^i\,_{jk}$ are defined by the relation \eqref{nablaofgamma}.
Then let us consider the covariant derivative
\begin{align*}
\langle\nabla\rangle_{e_j}e_k & =\frac{1}{vol({\bf
\hat{N}}_x)}\int_{\hat{\bf N}_x} \pi_2 (\nabla_{^he_j} \pi^*
e_k) \\
& =\,\frac{1}{vol({\hat{\bf N}}_x)}\int_{\hat{\bf N}_x}\pi_2
\Gamma^i\, _{jk}(x,y) \pi^* e_i \\
& =\frac{1}{vol(\hat{\bf N}_x)}\int_{\hat{\bf{N}}_x}\big(\Gamma^i\, _{jk}(x,y)\,\omega_x(y)\big)\,
 e_i.
\end{align*}
From the definition of the connection coefficients of a linear connection on {\bf M} one obtains
\begin{displaymath}
\langle\nabla\rangle_{e_j}e_k=\,^{\langle\nabla\rangle} \Gamma^i\,_{jk}(u) e_i (x)=\, \frac{1}{vol(\hat{\bf N}_x)}\int_{\hat{\bf N}_x}\big(\Gamma^i\, _{jk}(x,y)\,\omega_x(y)\big)\,e_i
\end{displaymath}
and the relation \eqref{gammapromediado} follows.
 \hfill$\Box$

The connection coefficients \eqref{gammapromediado} transforms as connection coefficients, as the following calculation shows,
\begin{align*}
\widetilde{\langle\Gamma\rangle}^i\,_{jk} (x)&=
\langle\tilde{\Gamma}^i\, _{jk}(x,y)\rangle\\
& = \frac{1}{\tilde{vol}(\hat{\bf N}_x)} \int_{{\bf
\hat{N}}_x}\,\, \tilde\Gamma^ i\,_{jk}(x,y)\,\omega_x(y)\\
& = \frac{1}{{vol}(\hat{\bf N}_x)} \int_{{\bf {N}}_x}\,\, \Big(\frac{\partial \tilde{x}^i}{\partial x^l}\frac{\partial x^p}{\partial \tilde{x}^j}\frac{\partial x^m}{\partial x^k}\Gamma^ l\,_{pm}(x,y)+\,\frac{\partial^2 x^b}{\partial \tilde{x}^j\partial \tilde{x}^k}\frac{\partial \tilde{x}^i}{\partial x^b}\Big)\,\omega_x(y)\\
& =\frac{\partial \tilde{x}^i}{\partial x^l}\frac{\partial x^p}{\partial \tilde{x}^j}\frac{\partial x^m}{\partial x^k}\frac{1}{{vol}(\hat{\bf N}_x)} \int_{{\hat{\bf N}}_x}\,\,\Gamma^ l\,_{pm}(x,y)\omega_x(y)+\,\frac{\partial^2 x^b}{\partial \tilde{x}^j\partial \tilde{x}^k}\frac{\partial \tilde{x}^i}{\partial x^b}\frac{1}{{vol}(\hat{\bf N}_x)}\int_{{\bf \hat{N}}_x} 1 \,\omega_x(y)\\
& =\, \frac{\partial \tilde{x}^i}{\partial x^l}\frac{\partial x^p}{\partial \tilde{x}^j}\frac{\partial x^m}{\partial x^k}\,\langle \Gamma^l\,_{pm}(x,y)\rangle +\, \frac{\partial^2 x^b}{\partial \tilde{x}^j\partial \tilde{x}^k}\frac{\partial \tilde{x}^i}{\partial x^b}\\
& =\, \frac{\partial \tilde{x}^i}{\partial x^l}\frac{\partial x^p}{\partial \tilde{x}^j}\frac{\partial x^m}{\partial x^k}\,\langle \Gamma\rangle^l\,_{pm}(x) +\, \frac{\partial^2 x^b}{\partial \tilde{x}^j\partial \tilde{x}^k}\frac{\partial \tilde{x}^i}{\partial x^b}.
\end{align*}

\subsection{Averaged Lorentz connection}
Let $({\bf M},\eta,F)$ be a Lorentzian Randers space and ${\bf N}$ the cone bundle of timelike vector fields respect to $\eta$.
Given the Lorentz connection $^L\nabla$ on the bundle $\pi^*|_{\bf N}{\bf
TM}\to {\bf N}$ as determined by {\it Proposition} \ref{proposiciononLnablaonpistartm}, there is an associated
averaged connection. The
measure is $\omega_x(y):=\,f(x,y)\, dvol(x,y)$, where $f(x,y)$ is a smooth, non-negative function with compact support on ${\bf N}$.
 The volume form is given by the $n$-form $(4.4)$.

Let us consider the following integrals along the fibers ${\bf N}_x$,
\begin{align*}
& {vol({\bf N}_x)}=\int _{{\bf N}_x}
f(x,y)\,dvol(x,y),\quad \langle y^i\rangle :=\frac{1}{vol({\bf N}_x)}\int _{{\bf N}_x} y^i
f(x,y)\,dvol(x,y),\\
& \langle y^m y^a y^l\rangle :=\frac{1}{vol({\bf N}_x)}\int _{{\bf N}_x} y^m y^a y^l
f(x,y)\,dvol(x,y).
\end{align*}
\begin{proposicion}
Let $({\bf M}, \eta, F)$ be a Lorentzian Randers space. Then
the averaged connection $\langle\,^L\nabla\rangle$ associated with the Lorentz connection $^L\nabla$ on the
pull-back bundle $\pi^*|_{{\bf N}}{\bf TM}\to {\bf N}$ is an affine,
torsion-free  connection on  ${\bf M}$.
\end{proposicion}
{\bf Proof.}
If we consider the average of the connection coefficients for the connection $^L\nabla$ on ${\pi}^*|_{\bf N}{\bf TM}$, one obtains the expression
\begin{align}
\langle\,^L\Gamma^i\,_{jk}(x,y)-\,^{\eta}\Gamma^i\,_{jk}(x)\rangle=\,\langle\big( L+\,T\big)^i\,_{jk}\rangle.
\label{Lnablagammaconnections}
\end{align}
This expression defines the connection coefficients of an affine connection respect to a particular frame by the relation \eqref{nablaofgammaaveraged}. It is indeed a symmetric connection, since the relation
\begin{align*}
\langle\,^L\Gamma^i\,_{kj}\rangle =\,\langle\,^L\Gamma^i\,_{kj}\rangle
\end{align*}
holds \hfill$\Box$

The connection $\langle \,^L\nabla \rangle$ is the averaged Lorentz connection. The expression \eqref{Lnablagammaconnections}  can be developed further. Taking into account the definition of $T$ and $L$ and that $^\eta\nabla$ is affine, we have
\begin{align}
\langle\,^L\Gamma^i\,_{jk}\rangle (x)=\,^\eta\Gamma^i\,_{jk}(x)+\,\langle\frac{1}{2\sqrt{\eta(y,y)}}\,\big( Sym(J(F^\sharp) (x)\otimes y^*)\rangle\,+J((y\cdot F)^\sharp)(x)\otimes\,
(\eta-\,\frac{1}{\eta(y,y)}\,y^*\otimes y^*)\big)^i\,_{jk}\rangle
\label{averageconnectionformula2}
\end{align}

\begin{corolario}
On the unit hyperboloid $\Sigma$, the connection coefficients of $\langle\,^L\nabla\rangle$ are given by
\begin{align}
\langle\,^L\Gamma^i\,_{jk}\rangle (x)=\,^\eta\Gamma^i\,_{jk}(x)+\,\langle \big( Sym(J(F^\sharp) (x)\otimes y^*)\rangle\,+\langle J((y\cdot F)^\sharp)(x)\otimes\,
(\eta-\,y^*\otimes y^*)\big)^i\,_{jk}\rangle
\label{gammalorentzpromediado}
\end{align}
with the indices $i,l,a,m=0,...,{n-1}$.
\end{corolario}
Equation \eqref{gammalorentzpromediado} can be developed further. Taking into account the linear properties of the averaging operation, we have the following relations,
\begin{align*}
& 1. \quad \langle Sym(J(F^\sharp) \otimes y^*)\rangle =\,Sym (J(F^\sharp) \otimes\langle  y^*\rangle ),\\
& 2. \quad \langle J((y\cdot F)^\sharp)\otimes\,\eta\rangle =\,(\langle y\rangle \cdot F)^\sharp\otimes\,\eta,\\
& 3. \quad \langle J((y\cdot F)^\sharp)\otimes\,\,y^*\otimes y^*\rangle =\,\langle y^*\otimes y^*\otimes y \rangle \cdot J(F^\sharp) ,
\end{align*}
where $\,\langle y^*\otimes y^*\otimes y \rangle \cdot J(F^\sharp) $ is the contraction of the tensor $\,\langle y^*\otimes y^*\otimes y \rangle $ with $J(F^\sharp) $ in the corresponding indices.

The following {\it Proposition} is direct from the properties of affine connections and the structure of the connection coefficients \eqref{gammalorentzpromediado},
\begin{proposicion} Let $({\bf M}, \eta, F)$ be a Lorentzian Randers space,
$f:{\bf N}\to {\bf R}$ a one particle distribution function with non-trivial compact support on ${\bf \Sigma}$
 and $\langle\,^L\nabla\rangle$ the averaged Lorentz connection. Then if $\eta(\dot{x},\dot{x})=1$, the following properties hold,
\begin{enumerate}

\item For each point $x\in {\bf M}$, there is a {\it normal coordinate
 system} such that the averaged coefficients are
zero.

\item Given a geodesic $\tilde{x}:{I}\to {\bf M}$ of $\langle\,^L\nabla\rangle$, there exits an adapted Fermi coordinate system to the geodesic.

\item For the Lorentz connection on ${\bf \Sigma}$, the averaged connection $\langle\,^L\nabla\rangle$ is determined by the first and
 third moments of the distribution function $f(x,y)$.
\end{enumerate}
\end{proposicion}
While the first and second properties are  general properties for averaged connection of linear connections on $\pi^*|_{\bf {N}}\hat{\bf N}$,
the third one is a specific property for the averaged Lorentz connection, since we are considering trajectories whose velocity fields are in the unit hyperboloid ${\bf N}$. This property has interesting applications in fluid dynamics and plasma physics \cite{GT1}.

\section{Comparison between the geodesics of the linear connections $\,^L\nabla$ and $\langle\,^L\nabla\rangle$}

In this section we compare the geodesic curves of  $\,^L\nabla$ and $\langle\,^L\nabla\rangle$. First, we introduce a metric structure in the space $\nabla_ {\bf N}$ of linear, torsion-free connections on $\pi^*|_{\bf N}{\bf TM}$. With such metric structure one can compare the linear connection $\,^L\nabla$ and the pull-back connection $\pi^*|_{\bf N}\langle\,^L\nabla\rangle$, since both are connections on $\pi^*|_{\bf N}{\bf TM}$. The comparison between the respective geodesic curves can be obtained from the comparison of the connections. As an application we show how  in the {\it ultra-relativistic limit}, the geodesics of $\,^L\nabla$ deviate from the geodesics of $\langle\,^L\nabla\rangle$, for the same initial conditions.
\subsection{A distance function on the space of torsion-free linear connections on $\pi^*|_{{\bf N}}{\bf TM}$}

Let us consider a time orientable Lorentzian manifold $({\bf M},\eta)$ and a  time orientation in $M$ is a smooth timelike vector field $U$. We will consider that $U$ is normalized.  In this case, one can consider the Euclidean metric $\bar{\eta}$, defined on {\bf M} by the expression
\begin{align}
\bar{\eta}(X,Y):=-\eta(X,Y)+2\eta(X,U)\eta(Y,U).
\label{metricbareta}
\end{align}
The restriction of $\bar{\eta}$ to ${\bf T}_x{\bf M}$ determines a scalar product on the
vector space ${\bf T}_x{\bf M}$, that in global coordinates $(y^0,...,y^{3})$ on ${\bf T}_x{\bf M}$ can be written as
\begin{align*}
\bar{\eta}_x=\bar{\eta}_{ij}(x,y) dy^i\otimes dy^j.
\end{align*}
Therefore, the pair $({\bf T}_x{\bf M},\bar{\eta}_x)$ is a finite dimensional Euclidean manifold and induces a Euclidean distance
function $d_{\bar{\eta}}$ on the tangent space ${\bf T}_x{\bf M}$,
 \begin{align*}
 d_{\bar{\eta}_x} :{\bf T}_x{\bf M}& \times {\bf T}_x{\bf M}\to {\bf R}\\
  & (y,z)\mapsto \bar{\eta}(y-z,y-z).
  \end{align*}
Let us assume that each of the functions
\begin{align*}
f_x:=f(x,\cdot)& :{\bf N}_x \to {\bf R},\quad (x,y)\mapsto f(x,y),\quad x\in {\bf M}
\end{align*}
has compact support on the cone ${\bf N}$ for each $x\in M$ and that the domain of $f(x,y)$ is compact on $M$.  The diameter of the distribution  $f_x(y)=f(x,y):{\bf T}_x{\bf M}\to {\bf R}$  is
\begin{align*}
\alpha_x:=supp\{d_{\bar{\eta}}(y_1,{y}_2)\, |\, y_1,{y}_2\in
support(f_x )\}
\end{align*}
and the {\it diameter of the distribution} $f$ is the parameter ${\alpha}:=supp \{{\alpha}_x,\, x\in {\bf M}\}\,\leq+\infty$.

The {\it normalized averaged vector field} is defined as
\begin{displaymath}
V(x)= \left\{
\begin{array}{l l}
  \frac{\langle\hat{y}\rangle }{\sqrt{\eta_{ij}\,\langle{y}^i\rangle\langle{y}^j\rangle}}\Big|_x,\, & \textrm{if }\eta(\langle y\rangle ,\langle y\rangle )\,> 0\\
  0, & \quad \textrm{otherwise}.\\
\end{array} \right.
\end{displaymath}
 $V(x)$ is not continuous on the boundary $\partial
\big(\pi(support(f))\big)$, where $\pi|_{\bf N}:{\bf N}\to {\bf R}$ is the canonical projection. In particular one has the relation
 \begin{displaymath}
 \|\langle y\rangle (x)\|^2_{\bar{\eta}}=\left\{
\begin{array}{l l}
 \bar{\eta}(\langle y\rangle ,\langle y\rangle )(x)\,> 1, & \quad (x,y)\in int(\pi(support(f)))\\
  0, & \quad (x,y)\in {\bf M}\setminus int(\pi(support(f))).\\
\end{array} \right.
 \end{displaymath}
Using partitions of the unity, one can approximate the main velocity vector field
$\langle y\rangle $ by another vector field  $^{\alpha^2}\langle y\rangle $ which is smooth
   in the whole spacetime {\bf M} and still timelike
   in an open sub-set of the interior of $\pi(support(f))$ (basically in $\pi(support(f))\setminus \partial \pi(support(f))$).
Therefore, there is an smooth extension of the vector field $V(x)$ to the whole $\pi(support(f))$,

Let us consider a measure, determined by collection  of $n$-forms $\{\omega_x(y)=f(x,y)\,dvol(x,y)\in\,T^*_xM,\,x\in {\bf M}\}$ such that each $\omega_x(y)$ are smooth on $y\in\,{\bf N}_x$. Then we have
\begin{proposicion}
Let $(M,\eta)$ be a Lorentzian spacetime and $\{\omega_x,\,x\in\,{\bf M}\}$ a measure with
$int(\pi(support(f))\neq \emptyset $. Then for $\epsilon< \frac{\alpha}{2}$, there is a $\mathcal{C}^\infty$
vector field $^{\epsilon}\langle y\rangle$ which coincides with the timelike vector field $\langle y\rangle$ in an open subset $^\epsilon\mathcal{O}$ of $\pi(support(f))$ and such the following relation holds:
 \begin{displaymath}
 \|\langle \hat{y}\rangle(x)-\,^{\epsilon}\langle\hat{y}\rangle(x)\|_{\bar{\eta}}\,=\mathcal{O}(\epsilon),\quad
 \forall \,x\in \,^{\epsilon}\mathcal{O}\cup {\bf M}\setminus {support(f)}.
\end{displaymath}
\label{renormalizedmainvelocity}
\end{proposicion}
{\bf Proof}. Let us consider an open subset $\mathcal{O}\,\subset int(\pi(support(f)))$
and the restriction of the vector field $\langle y\rangle$ to $\mathcal{O}$
denoted by $\langle y\rangle |_{\mathcal{O}}$. Then we consider the distance $d(z,\mathcal{O})$ between each point $z\in \partial \big(\pi support (f)\big)$ to the open
set $\mathcal{O}$,  using the Riemannian metric $\bar{\eta}$. Fixing $z$, this distance is realized by a point of the closure $w(z)\in\, \bar{\mathcal{O}}$. We can consider the geodesic segment between $z$ and $w$ respect to $\bar{\eta}$ for a open set $\mathcal{O}$ close
 to $int(\pi(support(f)))$ in such a way that
the geodesic segment exists. We call the maximal length of all those segments by $\epsilon$.
Then we can define the following interpolating vector field,
\begin{align}
v(x)=\left\{
\begin{array}{l l }
  0, & x \in {\bf M}\setminus int(support(\pi(f(x,y))),\\
  \frac{1}{d(z,\mathcal{O})}\,(1-\,d(\tilde{z},\mathcal{O}))v(w(z)), & x=\tilde{z}\in int(support(\pi(f(x,y)))\setminus \,\mathcal{O},\\
  v(x), & x\in\, \mathcal{O}.\\
\end{array} \right.
\label{renormalizedistance}
\end{align}
This vector field is continuous. One can find a smooth approximation using
bump functions based on the function $b(s)$,
\begin{align*}
b(s) = \left\{
\begin{array}{l l}
  e^{-\frac{k}{s}}, &  \quad s>0\\
  0, &  \quad s\leq 0\\
\end{array} \right.
\end{align*}
 in such a way that the resulting smooth
field is zero in ${\bf M}\setminus int(\pi(support(f)))$ (for a nice introduction to partition of the unity, see \cite{deRham} or \cite{Warner}). When it is positive, the variable $s$ is the distance function $d(z,\mathcal{O})$ and the maximal open set where \eqref{renormalizedistance} is smooth defines the open set $^\epsilon\mathcal{O}$.\hfill$\Box$

Therefore, if the diameter $\alpha$ of the distribution is small, $\epsilon$ can be of order $\alpha^2$ and the vector field $\langle y^i\rangle$ can be substituted by $^{\alpha^2}\langle y \rangle$ with an error of order $\alpha^2$. This fact implies that we can consider the vector field $\langle y \rangle$ as smooth, a fact that will be used in the following sub-sections, in particular in sub-section 3.3.

\subsection{Metric function on $\nabla_{\bf \Sigma^+}$}

Given a linear operator $A_x:{\bf T}_x{\bf M}\to {\bf T}_x{\bf M}$, its operator norm is defined by
\begin{align}
\|A\|_{\bar{\eta}}(x):=sup\,\Big\{\,\frac{\|A(y)\|_{\bar{\eta}}}{\|y\|_{\bar{\eta}}}
 (x),\,y\in{\bf T}_x{\bf M}\setminus \{0\}\,\Big\}.
 \label{normoperator}
\end{align}
Given two non-linear connections on {\bf TN}, one can define the corresponding linear connections on $\pi^*|_{{\bf \Sigma^+}}{\bf TM}$ by assuming the conditions in equation \eqref{structuralequationsforG}. This determines the respective horizontal lifts for each covariant derivative.
Let us consider the space $\nabla_{{\bf \Sigma^+}}$ of linear, torsion-free connections on $\pi^*|_{{\bf \Sigma^+}}{\bf TM}$ and define the function
\begin{align*}
\hat{d}_{\bar{\eta}}:\nabla_{{\bf \Sigma^+}}\times \,\nabla_{{\bf \Sigma^+}}\to {\bf R}
\end{align*}
\begin{align}
(\,^1\nabla,\,^2\nabla)\mapsto \sup\,\Big\{\frac{\sqrt{{\bar{\eta}}(x)(\,^1\nabla_{\iota(X)}
\pi^*|_{{\bf \Sigma^+}}X-\,^2\nabla_{\iota(X)} \pi^*|_{{\bf \Sigma^+}}X,\,^1\nabla_{\iota(X)} \pi_{{\bf \Sigma^+}}^*{X}-\,^2\nabla_{\iota(X)}
\pi^*|_{{\bf \Sigma^+}}X)}}{\bar{\eta}(X,X)},\,X\in\Gamma{\bf \Sigma^+}\Big\},
\label{distancefunctionforconnections}
\end{align}
with ${X}\in\,\Gamma{\bf \Sigma^+}$ and where $\iota(X)$ is the horizontal lift of $X$ using the non-linear connection associated with each of the geodesics sprays of each linear connection $^a\nabla$. Each of the horizontal lifts are done using different non-linear connection and therefore, are not necessarily the same horizontal lifts. However, we have that
\begin{proposicion}
The pair $(\nabla_{{\bf \Sigma^+}},\,\hat{d}_{\bar{\eta}})$ is a metric space.
\end{proposicion}
{\bf Proof}. The function \eqref{distancefunctionforconnections} is clearly symmetric. It is non-negative if $^1\nabla\neq \,^2\nabla$. The
distance between two arbitrary connections is zero iff
\begin{displaymath}
{\sqrt{{\bar{\eta}}(x)(\,^1\nabla_{\iota(X)}
\pi^*|_{{\bf \Sigma^+}}X-\,^2\nabla_{\iota(X)} \pi^*|_{{\bf \Sigma^+}}X,\,^1\nabla_{\iota(X)} \pi^*|_{{\bf \Sigma^+}}{X}-\,^2\nabla_{\iota(X)}
\pi^*|_{{\bf \Sigma^+}}X)}}=0
\end{displaymath}
for all $X\in {\bf \Sigma^+}_x$. The above equality holds iff
\begin{align*}
\,^1\nabla_{\iota(X)}
\pi^*|_{{\bf \Sigma^+}}X=\,^2\nabla_{\iota(X)} \pi^*|_{{\bf \Sigma^+}}X.
\end{align*}
Since the connections are torsion-free, this condition implies that $^1\nabla=\,^2\nabla$. Therefore, $\hat{d}_{\bar{\eta}}(\,^1\nabla,\,^2\nabla)=0$ iff $^1\nabla=\,^2\nabla$.
  The triangle inequality for $\hat{d}_{\bar{\eta}}$ follows from the triangle inequality for $\bar{\eta}$,
\begin{align*}
\hat{d}_{\bar{\eta}}(\nabla_1, \nabla_3) & =\,\sup\,\Big\{\frac{\sqrt{{\bar{\eta}}(x)(\,^1\nabla_{\iota(X)}
\pi^*|_{{\bf \Sigma^+}}X-\,^3\nabla_{\iota(X)} \pi^*|_{{\bf \Sigma^+}}X,\,^1\nabla_{\iota(X)} \pi^*|_{{\bf \Sigma^+}}{X}-\,^3\nabla_{\iota(X)}
\pi^*|_{{\bf \Sigma^+}}X)}}{\sqrt{\bar{\eta}(X,X)}}\Big\}\\
&\leq \,\sup\,\Big\{\frac{\sqrt{{\bar{\eta}}(x)(\,^1\nabla_{\iota(X)}
\pi^*|_{{\bf \Sigma^+}}X-\,^2\nabla_{\iota(X)} \pi^*|_{{\bf \Sigma^+}}X,\,^1\nabla_{\iota(X)} \pi^*|_{{\bf \Sigma^+}}{X}-\,^2\nabla_{\iota(X)}
\pi^*|_{{\bf \Sigma^+}}X)}}{\sqrt{\bar{\eta}(X,X)}}\Big\}\\
& +\,\sup\,\Big\{\frac{\sqrt{{\bar{\eta}}(x)(\,^2\nabla_{\iota(X)}
\pi^*|_{{\bf \Sigma^+}}X-\,^3\nabla_{\iota(X)} \pi^*|_{{\bf \Sigma^+}}X,\,^2\nabla_{\iota(X)} \pi^*|_{{\bf \Sigma^+}}{X}-\,^3\nabla_{\iota(X)}
\pi^*|_{{\bf \Sigma^+}}X)}}{\sqrt{\bar{\eta}(X,X)}}\Big\}\\
&\leq \hat{d}_{\bar{\eta}}(\,^1\nabla,\,^2\nabla)+\hat{d}_{\bar{\eta}}(\,^2\nabla,\,^3\nabla).
\end{align*}\hfill$\Box$

\subsection{Distance between the connections $^L\nabla$ and $\langle\,^L\nabla\rangle$}
Let us consider a Lorentzian Randers space $({\bf M},\eta, F)$.
Given the averaged connection $\langle\,^L\nabla\rangle$, we define an associated connection $\pi^*(\langle\,^L\nabla\rangle)$ on $\pi^*|_{{\bf \Sigma^+}}{\bf TM}$ such that the following conditions hold,
\begin{itemize}
\item If $Y\in \,\Gamma\,{\bf TM}$ and $\tilde{X} \in \Gamma {\bf T\Sigma}$ is an horizontal vector field, then
\begin{align*}
\pi^*|_{{\bf \Sigma^+}}(\langle\,^L\nabla\rangle)_{\tilde{X}}\,\pi^*|_{{\bf \Sigma^+}}Y :=\,\pi^*|_{{\bf \Sigma^+}} (\langle \,^L\nabla\rangle_{\pi_*(\tilde{X})}\,Y),
\end{align*}
where $\pi_*|_{{\bf \Sigma^+}}:{\bf TN}\to {\bf TM}$ is the differential of the projection $\pi|_{{\bf \Sigma^+}}:{\bf \Sigma^+}\to {\bf M}$.
\item If $V$ is a vertical tangent vector and $Z\in \,\Gamma \,{\bf T\Sigma}$, then
\begin{align*}
\pi^*|_{{\bf \Sigma^+}} \langle \,^L\nabla\rangle_V Z=0,
\end{align*}

\item For any function $f\in\,\mathcal{F}({\bf \Sigma^+})$ and $Z\in \,\Gamma \,{\bf TN},$ the covariant derivative is
\begin{align*}
 \pi^*|_{{\bf \Sigma^+}}(\langle\,^L\nabla\rangle )_{{Z}}\,f:=\,Z(f).
 \end{align*}
\end{itemize}
\begin{proposicion}
For the horizontal lift $^hX$, the following expression holds,
\begin{align}
\pi^*|_{{\bf \Sigma^+}}(\langle\,^L\nabla\rangle)_{\iota(X)}\,\pi^*|_{{\bf \Sigma^+}}Y =\,\pi^*|_{{\bf \Sigma^+}} (\langle \,^L\nabla\rangle_X\,Y).
\end{align}
\end{proposicion}
{\bf Proof}.
To show this, note that since the conditions of being torsion-free for $\pi^*|_{{\bf \Sigma^+}}(\langle\,^L\nabla\rangle)$, the derivatives along vertical directions are zero and therefore,
\begin{align*}
(\langle\,^L\nabla\rangle)_{\pi_*(\,\iota({X}))}=(\langle\,^L\nabla\rangle)_{(X)}.
\end{align*}\hfill$\Box$
\begin{proposicion}
The projection by $\pi:{\bf \Sigma^+}\to {\bf M}$ of the auto-parallel curves of $\pi^*|_{{\bf \Sigma^+}}\langle\,^L\nabla\rangle$ and  the auto-parallel curves $\langle\,^L\nabla\rangle$ coincide,
\begin{align*}
\pi^*|_{{\bf \Sigma^+}}\langle\,^L\nabla\rangle_{\iota(X)}X=0\,\Leftrightarrow\,\langle\,^L\nabla\rangle_{X}X=0.
\end{align*}
\end{proposicion}
{\bf Proof}. This is direct from the geodesic equations expressed in local coordinates.\hfill$\Box$

Therefore, although defined on different bundles, one compares the connections $^L\nabla$ and $\langle\,^L\nabla\rangle$ by comparing the connections $^L\nabla$ and $\pi|^*_{\bf \Sigma^+}(\langle\,^L\nabla\rangle)$. This is a legitim operation, since both are defined on the same bundle $\pi^*|_{{\bf \Sigma^+}} {\bf T M}$. Thus, one can define the distance between $\langle\,^L\nabla\rangle$ and $^L\nabla$ by the expression
\begin{align*}
\hat{d}_{\bar{\eta}}(\langle\,^L\nabla\rangle,\,^L\nabla):=\,\hat{d}_{\bar{\eta}}(\pi^*|_{{\bf \Sigma}^+}\langle\,^L\nabla\rangle,\,^L\nabla).
\end{align*}

\subsection*{Comparing $^L\nabla$ and $\langle\,^L\nabla\rangle$ in terms of the diameter of the distribution}
Let us consider the deviation vectors
\begin{align}
\delta(y):=\langle\hat{y}\rangle-y,\quad \delta(\hat{y}):=\langle\hat{y}\rangle-\hat{y}
\label{definitionofdelta}
\end{align}
where the {\it hat}-notation is used to distinguish integrated variables from the fixed coordinates $y^i$. In addition to the $\mathcal{F}({\bf \Sigma}^+)$-linearity of the average operation, there are two properties that we will use extensively in our calculations later,
\begin{align*}
\langle \delta(\hat{y})\rangle=0,\quad \langle\langle \hat{y}\rangle\rangle=\,\langle \hat{y}\rangle.
\end{align*}
\begin{proposicion}
Let $f(x,y)\in \,\mathcal{F}({\bf \Sigma}^+)$ be such that each function $f_x:{\bf \Sigma}^+_x\to {\bf M}$
 has compact and connected support with
$support(f_x)\subset{\bf \Sigma}^+_x$ and $int(\pi(support(f))\neq \,\emptyset$. Then the following relation holds,
\begin{align*}
(\,^L\nabla_y y-\,\pi^* |_{{\bf \Sigma}^+}\langle\,^L\nabla\rangle_y y)(x)=\,-\frac{1}{2}\pi^*|_{{\bf \Sigma}^+}\Big((\langle \delta(y)\rangle\cdot F)^\sharp \,(y\cdot\delta^*(y))+\, (\langle \hat(y)\rangle\cdot F)^\sharp(y\cdot\delta^*(y))^2))
\end{align*}
\begin{align}
+(y\cdot\langle\hat{y}\rangle^*)\big\langle(\delta(\hat{y})\cdot F)^\sharp\,(y\cdot\delta^*(\hat{y}))\big\rangle+\,\frac{1}{2}\langle (\hat{y}\cdot F)^\sharp \,(y\cdot\delta^*(\hat{y}))^2\rangle\Big).
\label{boundedistancebetweennablaandrhonabla}
\end{align}
\label{boundistancebetweenconnections}
\end{proposicion}
{\bf Proof}. From the expressions for $^L\nabla$ and $\langle\,^L\nabla\rangle$, we have the relation
\begin{align*}º
^L\nabla_y y - \, \pi^*|_{{\bf \Sigma}^+}\langle\,^L\nabla\rangle_y y =\,y\cdot\Big(y\cdot \big((T-\pi^*|_{{\bf \Sigma}^+}\langle T\rangle)+(L-\pi^*|_{{\bf \Sigma}^+}\langle L\rangle)\big)\Big),
\end{align*}
with the constraint $\eta(y,y)=1$ and $y$ future oriented.
To calculate the difference $y\cdot\big(y\cdot(T-\pi^*|_{{\bf \Sigma}^+}\langle T\rangle)\big)$ we proceed as follows. First, we note that the calculations of the difference can be make it pointwise, with $x\in {\bf M}$ and $y\in\,{\bf \Sigma}^+_x$, we can concentrate on evaluating the expressions
\begin{align*}
y\cdot\Big(y\cdot\big(L-\langle L\rangle\big)\Big),\quad
y\cdot\Big(y\cdot\big(T-\langle T\rangle\big)\Big).
\end{align*}
In the following calculations, we will use the following relations \eqref{definitionofdelta} several times.
For the longitudinal contributions, we have
\begin{align*}
\frac{1}{2}\,y\cdot\Big(y\cdot \big(L-\,\langle L\rangle\big)\Big) & =\,\frac{1}{2}\,y\cdot\Big(y\cdot \big(Sym\big( J(F^\sharp)  \otimes y^*\big)-\,Sym\big(J(F^\sharp)  \otimes \langle y\rangle^* \big)\Big)\\
& =\,\frac{1}{2}\,y\cdot\Big(y\cdot \big(Sym\big( J(F^\sharp)  \otimes (y-\langle y\rangle )^*  \big)\Big)\\
&  =\,\big( y\cdot F)^\sharp \,\big(y\cdot(y-\langle y\rangle )^*  \big)\\
&  =\,-\big( y\cdot F)^\sharp\, \big(y\cdot\delta^*(y)  \big).
\end{align*}
To evaluate the transverse contribution we make the following calculation,
\begin{align*}
\frac{1}{2}\,y\cdot\Big(y\cdot\big(T-\langle T\rangle\big)\Big) &=\,\frac{1}{2}\,y\cdot y\Big(\cdot J((y\cdot F)^\sharp)\otimes(\eta-\,y^*\otimes y^*)-\, \langle(\hat{y}\cdot F)^\sharp\otimes(\eta-\,\hat{y}^*\otimes \hat{y}^*)\rangle\Big)\\
& =\,\frac{1}{2}\Big(-(\langle \hat{y}\rangle\cdot F)^\sharp+\, \langle (\hat{y}\cdot F)^\sharp(y\cdot \hat{y}^*)^2\rangle \Big),
\end{align*}
where $\hat{y}$ stands for the {\it variable} which is  integrated in the average. The third term in the above expression can be developed as follows,
\begin{align*}
\langle (\hat{y}\cdot F)^\sharp(y\cdot \hat{y}^*)^2\rangle & =\langle (\hat{y}\cdot F)^\sharp(\langle (y\cdot (\langle \hat{y}\rangle-\,\delta (\hat{y}))^*)^2\rangle
\end{align*}
Thus, developing the square,
\begin{align}
 \langle (\hat{y}\cdot F)^\sharp(y\cdot \hat{y}^*)^2\rangle=\,\big\langle (\hat{y}\cdot F)^\sharp \,\Big( (y\cdot\langle \hat{y}\rangle^*)^2-\,2\,(y\cdot\langle\hat{y}\rangle^*)(y\cdot\delta^*(\hat{y}))+\,(y\cdot\delta^*(\hat{y}))^2\Big)\big\rangle
\label{desarrollott}
\end{align}
Each of the above individual terms can be developed further as follows.

The first term  in \eqref{desarrollott} is equivalent to
\begin{align*}
\langle (\hat{y}\cdot F)^\sharp \,\big(y\cdot \langle \hat{y} \rangle^*)^2\rangle & =\,  \langle (\hat{y}\cdot F)^\sharp \,\big(y\cdot(y^* +\delta^*(y))\big)^2\rangle\\
& = \,(\langle \hat{y}\rangle\cdot F)^\sharp \,\Big(1+2\,(y\cdot\delta^*(y))+\, (y\cdot\delta^*(y))^2\Big),
\end{align*}
where in the last equality the linearity of the average operation $\langle\cdot\rangle$ and the normalization condition $y\cdot y^*=\eta(y,y)=1$ has been used.

For the second term in \eqref{desarrollott} we have
\begin{align*}
-2\,\langle (\hat{y}\cdot F)^\sharp \,(y\cdot\langle\hat{y}\rangle^*)(y\cdot\delta^*(\hat{y}))\rangle & =\,-2(y\cdot\langle\hat{y}\rangle^*)\,\big\langle (\hat{y}\cdot F)^\sharp (y\cdot\delta^*(\hat{y}))\big\rangle\\
& =\,-2(y\cdot\langle\hat{y}\rangle^*)\big\langle (\langle\hat{y}\rangle-\,\delta(\hat{y}))\cdot F)^\sharp\,(y\cdot\delta^*(\hat{y})\big\rangle\\
& = \,-2(y\cdot\langle\hat{y}\rangle^*)\big\langle (\langle\hat{y}\rangle\cdot F)^\sharp\,(y\cdot\delta^*(\hat{y})\big\rangle\,+2(y\cdot\langle\hat{y}\rangle^*)(\delta(\hat{y})\cdot F)^\sharp\,(y\cdot\delta^*(\hat{y}))\big\rangle\\
& =\,-2(y\cdot\langle\hat{y}\rangle^*)\,(\langle\hat{y}\rangle\cdot F)^\sharp\,\big\langle y\cdot\delta^*(\hat{y})\big\rangle\,+2(y\cdot\langle\hat{y}\rangle^*)(\delta(\hat{y})\cdot F)^\sharp\,\langle(y\cdot\delta^*(\hat{y}))\rangle\\
& =\,2(y\cdot\langle\hat{y}\rangle^*)(\delta(\hat{y})\cdot F)^\sharp\,(y\cdot\delta^*(\hat{y}))\big\rangle,
\end{align*}
where in the last equality we have use the fact that $\langle y\cdot\delta^*(\hat{y})\rangle=\,y\cdot\langle \delta(\hat{y})\rangle$ and also $\langle \delta(\hat{y})\rangle=0$.

The third term in \eqref{desarrollott} is of second order in $\delta(y)$ already.

Writing all together the contributions to $\frac{1}{2}y\cdot(T-\langle T\rangle)$ we obtain
\begin{align*}
\frac{1}{2}y\cdot(T-\langle T\rangle) & =\,\,(\langle \hat{y}\rangle\cdot F)^\sharp \,\big(\,(y\cdot\delta^*(y))+\, (y\cdot\delta^*(y))^2)\big)\\
& +(y\cdot\langle\hat{y}\rangle^*)\big\langle(\delta(\hat{y})\cdot F)^\sharp\,(y\cdot\delta^*(\hat{y}))\big\rangle+\,\frac{1}{2}\langle (\hat{y}\cdot F)^\sharp \,(y\cdot\delta^*(\hat{y}))^2\rangle.
\end{align*}

Thus, writing all these contributions together and after a short simplification, one obtains
\begin{align*}
\frac{1}{2}y\cdot\Big(y\cdot\big(L-\langle L\rangle\big)+\,\big(T-\langle T\rangle\big)\Big)& =\,(\langle \delta(y)\rangle\cdot F)^\sharp \,(y\cdot\delta^*(y))+\, (\langle \hat(y)\rangle\cdot F)^\sharp(y\cdot\delta^*(y))^2))\\
& +(y\cdot\langle\hat{y}\rangle^*)\big\langle(\delta(\hat{y})\cdot F)^\sharp\,(y\cdot\delta^*(\hat{y}))\big\rangle+\,\frac{1}{2}\langle (\hat{y}\cdot F)^\sharp \,(y\cdot\delta^*(\hat{y}))^2\rangle.
\end{align*}
This relation implies the formula \eqref{boundedistancebetweennablaandrhonabla}.
\hfill$\Box$
\begin{proposicion}
Let $({\bf M},\eta, F)$ be a Lorentzian Randers space and $^L\nabla$ the associated Lorentz connection. For each fixed $x\in{\bf M}$, $int(\pi(support(f)))\neq \, \emptyset$, let
$f_x:{\bf \Sigma}^+_x\to {\bf R}$ to have compact and connected support
such that $\alpha:=\,sup \{\alpha_x,\,\,x\in {\bf M}\}<<1$. Then the following expression holds:
\begin{align}
\hat{d}_{\bar{\eta}}(\,^L\nabla,\langle\,^L\nabla\rangle )(x)\leq \,\| F) \|_{\bar{\eta}}(x)\Big(\frac{3}{2}\|\langle \hat{y}\rangle\|_{\bar{\eta}}+1\Big)(2\alpha+\alpha^2)^2.
\label{boundformula}
\end{align}
\label{distancebetweenconnections}
\end{proposicion}
{\bf Proof}. Let us consider the norm defined from $\bar{\eta}$ instead of the norm defined by $^{\epsilon}\bar{\eta}$. This substitution can be done because the differences between the metrics $\bar{\eta}$ and $^\epsilon\bar{\eta}$ are small (at least of order $\alpha$, by relating the parameter $k$ with $\alpha$ conveniently).
From equation \eqref{boundedistancebetweennablaandrhonabla} one obtains
\begin{align*}
\|\, ^L\nabla_y y-\, \pi^*\langle\, ^L\nabla\rangle _y y\|_{\bar{\eta}} & =\| \frac{1}{2}y\cdot\Big(y\cdot\big(L-\langle L\rangle\big)+\,\big(T-\langle T\rangle\big)\Big)\|_{\bar{\eta}}\\
 & =\,\|(\langle \delta(y)\rangle\cdot F)^\sharp \,(y\cdot\delta^*(y))+\, (\langle \hat(y)\rangle\cdot F)^\sharp(y\cdot\delta^*(y))^2))\\
& +(y\cdot\langle\hat{y}\rangle^*)\big\langle(\delta(\hat{y})\cdot F)^\sharp\,(y\cdot\delta^*(\hat{y}))\big\rangle+\,\frac{1}{2}\langle (\hat{y}\cdot F)^\sharp \,(y\cdot\delta^*(\hat{y}))^2\rangle\|_{\bar{\eta}}\\
&\leq \,\|(\langle \delta(y)\rangle\cdot F)^\sharp \,(y\cdot\delta^*(y))\|_{\bar{\eta}}+\, \|(\langle \hat(y)\rangle\cdot F)^\sharp(y\cdot\delta^*(y))^2))\|_{\bar{\eta}}\\
& +\|(y\cdot\langle\hat{y}\rangle^*)\big\langle(\delta(\hat{y})\cdot F)^\sharp\,(y\cdot\delta^*(\hat{y}))\big\rangle\|_{\bar{\eta}}+\,\|\frac{1}{2}\langle (\hat{y}\cdot F)^\sharp \,(y\cdot\delta^*(\hat{y}))^2\rangle\|_{\bar{\eta}}.
\end{align*}
Each of these four terms can be bound individually. To do this, we need some geometric inequalities first. Since the support of the distribution function $f(x,y)$ is compact and connected, one can write the decomposition $\langle\hat{y}\rangle(x)=\epsilon (x)+z(x)$
 with the property that $z(x)\in support(f_x)$ and the norm of $\epsilon(x)$ is bounded by $\alpha$, using the metric $\bar{\eta}$.
 In the case of the Lorentzian metric
 one can check by geometric inspection that $\|\epsilon(x)\|_{\bar{\eta}}
 \leq \,\alpha$.
First,
we have the following geometric bound
\begin{align*}
\|\delta(x,y)\|_{\bar{\eta}}\,& \leq\|\langle\hat{y}\rangle (x)\,-y\|_{\bar{\eta}}\leq \|\epsilon+{z}(x)-y\|_{\bar{\eta}} \leq \|\epsilon\|_{\bar{\eta}}+\|z(x)-y\|_{\bar{\eta}}\,\leq 2 \alpha+\alpha^2.
\end{align*}
To prove this bound, we use the relation $\delta(x,{y})=\,\langle{y}\rangle -{y}$, one finds also the following bound,
\begin{align*}
|y\cdot\delta^*(y)| & =|y\cdot\langle\hat{y}\rangle^* -1| =|\langle\hat{y}\rangle\cdot(y-\,\langle\hat{y}\rangle \,+\langle\hat{y}\rangle)^*\,-1| \leq |\langle\hat{y}\rangle \cdot(y\,-\langle\hat{y}\rangle)^*| +|\langle \hat{y}\rangle\cdot\langle\hat{y}\rangle^* -1|.
\end{align*}

Using the Cauchy-Schwartz inequality for $\bar{\eta}$, we obtain from the above expression
\begin{align*}
&|\delta \cdot y^*| \leq \, \|\langle\hat{y}\rangle \|_{\bar{\eta}}\,\|(y\,-\langle\hat{y}\rangle )\|_{\bar{\eta}}+|(\langle\hat{y}\rangle \cdot\langle \hat{y}\rangle^*) -1|\leq \|\langle \hat{y}\rangle\|_{\bar{\eta}}\,{\alpha}\,+|(\langle\hat{y}\rangle\cdot\langle\hat{y}\rangle^*)\,-1|\\
&\leq \sqrt{1+\,\|\epsilon\|_{\bar{\eta}}}\,\alpha\,+(\sqrt{1+\,\|\epsilon\|_{\bar{\eta}}}-1)\leq \sqrt{1+\,{\alpha}}\,\alpha\,+(\sqrt{1+\,{\alpha}}-1)\leq 2\alpha\,+{\alpha^2}.
\end{align*}
There are two further inequalities, direct consequence from the definition of a norm of a operator \eqref{normoperator},
\begin{align*}
\|(\delta(y)\cdot F)^\sharp\|_{{\bar\eta}}\leq \,\|(\delta(y)\|_{{\bar\eta}}\,\|F\|_{{\bar\eta}}\leq\alpha\,\|F\|_{{\bar\eta}},\quad
\|(\langle\hat{y}\rangle\cdot F)^\sharp\|_{{\bar\eta}}\leq\,\|\langle\hat{y}\rangle\|_{{\bar\eta}}\,\|F\|_{{\bar\eta}}.
\end{align*}
When apply to bound each of the four contributions to the difference $\|\, ^L\nabla_y y-\, \pi^*\langle\, ^L\nabla\rangle _y y\|_{\bar{\eta}} $ we obtain, after some re-arrangement, the right side of the expression \eqref{boundformula}.

\hfill$\Box$

\subsection{Comparison of the geodesics of $^L\nabla$ and $\langle\,^L\nabla\rangle$}

Since the dynamics of the Lorentz force equation and the averaged Lorentz force equation are determined by the respective connections and we have shown that such connections {\it differ} by a second or higher order factor in $\alpha$, it is not a surprise that the auto-parallel curves also differ in terms of polynomials in $\alpha$.

There are two vector fields  playing a relevant role in the following considerations. One is the vector field $U_{lab}$ associated with an observer in the laboratory. Its integral curves correspond with the world-line of the laboratory observers. The second vector field is the averaged energy-momentum vector field $\langle y\rangle$. When it differs from zero, the integral curves of $\langle y\rangle$ corresponds to the time evolution of the averaged energy-momentum vector of the distribution function $f(x,y)$ (up to order $\alpha^2$ \cite{GT2}).

Given $x\in \,{\bf M}$ such that $\langle y\rangle$ is not zero, there is an instantaneous Lorentz boost transformation from a coordinate system adapted to the vector field $U_{lab}$ to a coordinate system adapted to the vector field $U(x)$.

The energy function $E$ of the distribution function $f$ is defined to be the real function
\begin{equation}
E:{\bf M}\to {\bf R},\quad
x\mapsto E(x):=\,\inf\{ y^0,\, y\in support(f_x)\},
\end{equation}
where $y^0$ is the $0$-component of the tangent velocity vector,
 measured in the laboratory coordinate frame associated with $U_{lab}$.
Thus, let us fix a particular solution $x:[0,T]\to M$ of the Lorentz force equation, with $t$ being the time measured by a clock associated with the laboratory coordinate system $U_{lab}$. Then ${\gamma}(t)$ is the gamma factor of the Lorentz boost from the laboratory frame to the co-moving frame associated with $x:[0,T]\to {\bf M}$. Given an initial condition for the Lorentz force equation, the gamma factor ${\gamma}\sim \,E$ is a function of the time $t$. Similarly, one can consider the solutions of the averaged Lorentz force equation and the corresponding gamma factors $\tilde{\gamma}$. We will consider that the domain of the field $F$ is compact. Thus, being an external field, $F_{\max}:=\max\{\|F\|_{\hat{\eta}},\,x\in {\bf M}\}<\,\infty$.
\begin{teorema}
Let $({\bf M},\eta, F)$ be a Lorentzian Randers space. Let us assume that
\begin{itemize}

\item The ultra-relativistic limit holds: the energy function of the beam $E(x(t))$ is much larger than the rest mass of the particles, $E(x(t))>>1$.

\item The change in the energy function is adiabatic in the sense that $\frac{d}{dt} log E<<1$.
\end{itemize}
Then for the same arbitrary initial condition $(x(0),\dot{x}(0))$, there is a common time interval $[0,T]$ where the solutions of the ordinary differential equations
$^L\nabla_{\dot{x}} \dot{x}=0$ and $ \langle\,^L\nabla\rangle_{\dot{\tilde{x}}} \dot{\tilde{x}}=0$ exist, are unique and
differ in such a way that
\begin{align}
|\tilde{x}^i(t)-\, x^i(t)|\leq\, F_{\max}\Big(\frac{3}{2}\|\langle \hat{y}\rangle\|_{\bar{\eta}}+1\Big)(2\alpha+\alpha^2)^2\,E^{-2}(x)\,t^2,\quad i=0,...,n-1.
\label{formuladiferenciadeposicion}
\end{align}
\label{teoremareferenciapositions}
\end{teorema}
{\bf Proof}. Existence and uniqueness follow from standard ODE theory \cite{Chicone}. We calculate the distance measured in the laboratory frame
between $x(t)$ and $\tilde{x}(t)$, solutions of the geodesic equations of the corresponding
 connections $^L \nabla$ and $\langle\,^L\nabla\rangle$, when both geodesics have the same initial
 conditions $(x(0),\dot{x}(0))$. In order to do this we use the formal solution of a second order differential equation, writing the
 solution of the Lorentz force equation $^L\nabla_{\dot{x}} \dot{x}=0$ as
\begin{align}
x^i (t)=x^i (0)+\int^t _0 ds \Big(\dot{x}^i(0) + \int^s _0
dl\,\ddot{x}^i(l)\Big),\quad i=0,...,n-1.
\label{formalsolutionforx}
\end{align}
Since the initial conditions for both connections are the same, the analogous
relation for the geodesics of the averaged connection $\langle\,^L\nabla\rangle$ is
\begin{align}
\tilde{x}^i (t)=x^i (0)+\int^t _0 ds \Big(\dot{x}^i(0) + \int^s _0
dl\ddot{\tilde{x}}^i(l)\Big),\quad i=0,...,n-1.
\label{formalsolutionfortildex}
\end{align}
The relations between proper times and
coordinate time in the laboratory frame are
\begin{align*}
d\tau=\gamma^{-1}dt, \quad d\tilde{\tau}=\tilde{\gamma}^{-1}dt.
\end{align*}
This implies the following relations
\begin{align*}
\frac{d}{dt}=\gamma^{-1}\frac{d}{d\tau},\quad
\frac{d}{dt}=\tilde{\gamma}^{-1}\frac{d}{d\tilde{\tau}}.
\end{align*}
We can assume the approximation $\tilde{\gamma}=\gamma$ up to order $\alpha^2$.
This is because the distance between the  connections $^L\nabla$ and $\langle\,^L\nabla\rangle$ is at least of order  $\alpha^2$ and therefore, $\gamma$ (respectively ($\tilde{\gamma}$) has a smooth dependence on the connection coefficients $^L\nabla^i_{jk}$ (respectively $\langle\,^L\nabla^i_{jk}\rangle$). Then by the formal solutions \eqref{formalsolutionforx} and \eqref{formalsolutionfortildex} one obtains the expression
\begin{align*}
|\tilde{x}^i(t)-{x}^i(t)|\leq \, t\int^t _0 dl E^{-2}\,\|
\frac{d^2\tilde{x}(l)}{dl^2}-\frac{d^2{x}(l)}{dl^2}\|_{\bar{\eta}},\quad i=0,...,n-1.
\end{align*}
Note that by the definition \eqref{distancefunctionforconnections}, one has the relation
\begin{align*}
\|\frac{d^2\tilde{x}(l)}{dl^2}-\frac{d^2{x}(l)}{dl^2}\|_{\bar{\eta}}\leq \,d_{\bar{\eta}} (\, ^L\nabla,\, \pi^*\langle\,^L\nabla\rangle ),\quad i=0,...,n-1.
\end{align*}
As a result, in the ultra-relativistic regime we have that
\begin{align*}
|\tilde{x}^i(t)-{x}^i(t)|\leq\, \alpha\,E^{-2}
\,\cdot d_{\bar{\eta}} (\, ^L\nabla,\, \pi^*\langle\,^L\nabla\rangle )\, t^2,\quad i=0,...,n-1.
\end{align*}

In a similar way, we can compare the velocity vector of the curves $\tilde{x}$ and $x$,
\begin{teorema} Under the same hypothesis as in {\it Theorem} \ref{teoremareferenciapositions}, the difference between the velocity vectors of the respective geodesics of $^L\nabla$ and $\langle\,^L\nabla\rangle$ is given by
\begin{align}
|\dot{\tilde{x}}^i(t)-\dot{x}^i(t)|\leq F_{\max}\Big(\frac{3}{2}\|\langle \hat{y}\rangle\|_{\bar{\eta}}+1\Big)(2\alpha+\alpha^2)^2\,E^{-1}\,t,\quad i=0,...,n-1.
\label{formuladiferenciadevelocidad}
\end{align}
\label{teoremareferenciaspeeds}
\end{teorema}
{\bf Proof}. The proof is analogous to the proof of {\it Theorem} \ref{teoremareferenciapositions},
 although based on the following formula for the tangent velocity field along a curve:
\begin{equation}
\dot{x}^i(t)=\dot{x}^i(0)+\int^t_0\ddot{x}^i(l)dl,\quad i=0,...,n-1.
\end{equation}\hfill$\Box$

\subsection{Examples}
There are at least two situations where {\it Theorem} \ref{teoremareferenciapositions} and {\it Theorem} \ref{teoremareferenciaspeeds} apply in a natural way:
\begin{enumerate}
\item The {\it ultra-relativistic regime} is defined as the dynamical regime such that in the laboratory frame, the energy of each particle in the bunch is considered asymptotically as
$E\to \,+\infty$, keeping $\alpha$ bounded.
 In this regime, for the dynamics under the hypothesis of {\it Theorems} \ref{teoremareferenciapositions} and \ref{teoremareferenciaspeeds} one has the limits
 \begin{align}
|x^i(t)-\,\tilde{x}^i(t)|\to 0, \quad |\dot{x}^i(t)-\,\dot{\tilde{x}}^i(t)|\to 0,\quad i=0,...,n-1,
 \label{limitboundeinfty}
 \end{align}
for distribution functions $f(x,y)$ with  $\alpha>0$. The ultra-relativistic limit bound \eqref{limitboundeinfty} is also valid not only for short times, but also for long time evolutions.

\item For the Dirac delta distribution, the limit  where {\it Theorems} \ref{teoremareferenciapositions} and \ref{teoremareferenciaspeeds} provides an exact approximation scheme.
\begin{align}
f(x,y)=\delta(y-V(x))\Psi(x),\quad x\in {\bf M}.
\end{align}
Dirac delta distribution function corresponds to the charged cold fluid model.
Since the width in the space of tangent velocities of the distribution is ${\alpha}=0$, and one has that
\begin{align}
\lim_{\alpha\to 0} |x^i(t)-\,\tilde{x}^i(t)|=0,\quad \lim_{\alpha\to 0} |\dot{x}^i(t)-\,\dot{\tilde{x}}^i(t)|= 0,\quad i=0,...,n-1,
\label{limitalphao}
\end{align}
\end{enumerate}
Deviations from the above two limit cases imply deviations in the approximation of the Lorentz dynamics by the Lorentz force dynamics.  Such deviations can be handle perturbatively in terms of polynomial functions in $E^{-1}$ and $\alpha$.

\subsection{Limits on the applicability of the approximation by the averaged model}
Let us consider the formula \eqref{formuladiferenciadeposicion}. At leading term in $E$ and $\alpha$, the differences of coordinates is given by
\begin{displaymath}
|x^i(t)-\,\tilde{x}^i(t)|\leq\, F_{\max}\Big(\frac{3}{2}\|\langle \hat{y}\rangle\|_{\bar{\eta}}+1\Big)(2\alpha+\alpha^2)^2\,E^{-2}(x)\,t^2\quad i=0,...,n-1.
\end{displaymath}
There are two different contributions to the bound in the right side of this expression. The dominant one is proportional to $\|\langle \hat{y}\rangle\|_{\bar{\eta}}$, since in the ultra-relativistic limit, it is not bounded. A natural way to bound this factor is to allow the product $\alpha\|\langle \hat{y}\rangle\|_{\bar{\eta}}$ uniformly bounded by a finite constant $C$. The argument of why this bound happens is that there are limitations on the stability of the system: reaching large enough spread $\alpha$ at a very high energy is unstable (for instance, in a  particle accelerator, the bunch become  un-stable and decay fast or the beam reach the physical bounds of the pipe. Thus, this constraint is natural. Thus we impose the condition
\begin{align}
\langle \hat{y}\rangle( t) \alpha(t)\,\leq C, \quad \forall \,t\in I,
\end{align}
where $C$ is a constant. Therefore, the bounds are
\begin{align}
|x^i(t)-\,\tilde{x}^i(t)|\leq\,C F_{\max}\Big(4\|\langle \hat{y}\rangle\|_{\bar{\eta}} \alpha+\mathcal{O}(\alpha^2)\Big)E^{-2}(x)\,t^2,\quad i=0,...,n-1.
\end{align}
and
\begin{align}
|\dot{x}^i(t)-\,\dot{\tilde{x}}^i(t)|\leq\,C F_{\max}\Big(4\|\langle \hat{y}\rangle\|_{\bar{\eta}} \alpha+\mathcal{O}(\alpha^2)\Big)E^{-1}(x)\,t,\quad i=0,...,n-1.
\end{align}

In a the application of the theory to the dynamics of a bunch of charged particles in a particle accelerator beam, a natural maximal distance $L_0$ between two particles of the same bunch appears.
For instance, in an accelerator machine, let $L_0$ be the {\it effective diameter} of the accelerator pipe.
In such situation the averaged model is not applicable when  $|{\tilde{x}}^i(t)-{x}^i(t)|$ is of the order of $L_0$ or larger. This can happen after a time evolution such that the difference $\|{\tilde{x}}^i(t)-{x}^i(t)\|=L_0$.
The characteristic time where the averaged model loses validity is
\begin{align}
t_{max}\sim \, E\,\Big(\,\frac{L_0}{C\,{\alpha} F_{\max}}\Big)^{\frac{1}{2}}.
\label{tiempocaracteristico1}
\end{align}
Reasoning in a similar way with the difference in velocities given by formula \eqref{formuladiferenciadevelocidad}, there is a second maximal critical time for the averaged model can start to fail. This corresponds when the difference $|\dot{{\tilde{x}}}^i(t)-\dot{x}^i(t)|$ is of order of the speed of light $c=1$. In this case, the characteristic time is
\begin{align}
t' _{max}\sim \, E\,\Big(\,\frac{c}{ {C\,\alpha}{F}_{\max}}\Big),
\label{tiempocaracteristico2}
\end{align}
where $c$ is the speed of light.
The limit of applicability of the averaged model is given by $t_{\max}$. In general, it will be given by $\min\{t_{\max},t'_{\max}\}$.

From the above discussion one has the following results,
\begin{corolario}
In the double limit $E\rightarrow \,+\infty$ and ${\alpha}\rightarrow 0$ and for distributions such that $\frac{d}{dt} log E<<1$ and $|\langle \hat{y}\rangle( t) \alpha(t)|\,\leq C$, one has that $t_{\max}\to \,+ \infty$.
\end{corolario}

{\small\section{Applications of the averaged Lorentz dynamics}}

Some physical applications of the averaged Lorentz force equation are highlighted below:
\begin{itemize}

\item {\bf Definition of reference trajectory in beam dynamics}. The solutions of the averaged Lorentz force equation can be used as the {\it reference trajectory} in
beam particle dynamics \cite{W}. This application is motivated by the following properties,
\begin{enumerate}
\item By {\it Theorem}  \ref{teoremareferenciapositions} and {\it Theorem} \ref{teoremareferenciaspeeds}, for the same initial
conditions, in the ultra-relativistic limit and for narrow
probability distributions, the difference between the original Lorentz
trajectory and the averaged Lorentz trajectory is small.

\item For narrow distributions in the ultra-relativistic limit, the integral curves of the averaged velocity field coincide at least up to order ${\alpha}$  with the integral curves of the auto-parallel vector field of the averaged connection  \cite{GT1}.

\end{enumerate}
The reference trajectory is a fundamental notion in beam dynamics. However, given a particular reference trajectory, it does not necessarily coincide with the trajectory of a real particle of the bunch and it can also happen that it is not directly observable. Therefore, we suggest that the reference trajectory should be defined as a solution of the averaged Lorentz dynamics. This is based on the fact that the solutions of the averaged Lorentz equation are {\it observable trajectories}, corresponding (up to order $\alpha$) to the evolution of the averaged moment $\langle y\rangle$. This idea has been further developed in \cite{Ricardo012}.

\item {\bf Applications in fluid dynamics}. The averaged dynamics has been used in an alternative derivation of relativistic fluid models from relativistic kinetic model \cite{GT2,GT1}. The motivation for the new derivation is based on the following two points,
    \begin{enumerate}
   \item  Fluid models emerge from kinetic theory as an averaged description of kinetic models \cite{Amendt, D, PenniAnile}. The standard proofs of these derivations are based on assumptions on the higher order moments of the distribution function. These assumptions are excluded from direct experimental control, in the very extreme conditions of bunches of particles in an accelerator. Using the averaged dynamics, we have approached the problem without making hypothesis on the higher moments: only moments up to order three appear in our considerations \cite{GT2,GT1}.

   \item Since the averaged Lorentz connection is affine and symmetric,
 one can make use of normal coordinates or Fermi coordinates. This allowed us to simplify some technical calculations that otherwise must be done with additional hypothesis on the distribution function \cite{GT1}.
   \end{enumerate}
 Since the averaged Lorentz connection is affine and symmetric, in the case of irreducible connections, the possible holonomy is classified by Merkulov-Schachhoefer \cite{MerkulovSchachhoefer}. Although the holonomy classification applies to the general averaged connection, in the case of the averaged Lorentz connection, there is a strong interplay between the two form $F$, the distribution function $f$ and the possible holonomy representation. In addition, if the distribution function $f(x,y)$ is determined by the $2$-form $F$ (for instance, via solutions of the Vlasov equation \cite{Ehlers}), then one can argue that the holonomy group of the affine connection $\langle\,^L\nabla\rangle$ is determined by $F$ as well, that is, by the second class of the de Rham cohomology group $H^2_{dR}({\bf M},{\bf R}).$ This is a further constraint on the holonomy group of the connection $\langle\,^L\nabla\rangle$. and it could have interesting applications  in charged plasma dynamics, specially in classification of solutions.

\item {\bf Universality of emergent fluid models}. In addition to the ultra-relativistic regime, the approximation of the Lorentz dynamics by the averaged Lorentz dynamics requires narrowness of the distribution function $f(x,y)$. In particular, the averaged model only requires information of the lower moments of the distribution $f(x,y)$ and is insensitive to higher moments. This implies the following
\begin{proposicion}
In the ultra-relativistic limit and for narrow distribution functions $f(x,y)$ as specified in {\it Theorems} \ref{teoremareferenciapositions}, the averaged dynamics depends on the zero, first, second and third moments.
\end{proposicion}
 Thus, one can assume that moments of higher order than three vanish, providing an argument for the universality of fluid models.

\item {\bf Extension to chromo-hydrodynamics}.
 The possibly extension to non-abelian theories is non-trivial, since the isotropy spin equation is first order \cite{Wong}. However, one can write the Yang-Mills equations as geodesic equations \cite{Kerner1968, Montgomery}. Therefore, the averaged method becomes applicable. This can have potential applications in chromo-hydrodynamics \cite{GibbonsHolmKupershmidt}.

 \end{itemize}

\section{Discussion}

The averaging method for connections was introduced  in the context of Finsler geometry \cite{Ricardo}. It is a generalization of the {\it averaging along the fiber operation} of classical mechanics \cite{Arnold}
and algebraic topology \cite{BottTu}. Thus, the common topological properties encoded in the averaged and non-averaged geometric objects are highlighted  by the property of {\it convex invariance} \cite{Ricardo}. In this paper, the averaging method was applied to the dynamics of point charged particles described by the Lorentz force equation. We have shown that in the ultra-relativistic limit and for narrow  distributions, the Lorentz connection $^L\nabla$ can be substituted by the averaged connection $\langle\,^L\nabla\rangle$, going beyond the common topological properties of both models. This happens for the connection $^L\nabla$ as determined by {\it proposition} \ref{proposiciononLnablaonpistartm}.

The hypothesis in {\it Theorem} \ref{teoremareferenciapositions} and {\it Theorem} \ref{teoremareferenciaspeeds} can be motivated by physical considerations:
\begin{itemize}

\item Ultra-relativistic dynamics means that the minimal value of the component $y^0$
on the support of the distribution function $f(x,y)$ is much larger than the rest mass of the particles  $m=1$. The definition of ultra-relativistic
limit is a gauge invariant notion.

\item The physical interpretation for the narrowness condition $\alpha<<1$ is clear: the diameter of the
distribution is much smaller than the rest mass of each of the particles in the bunch. The condition of narrowness
is gauge and Lorentz invariant (since one has choose first the laboratory coordinate frame).
We can compare our definition of narrowness with the {\it invariant warm fluid condition} \cite{Amendt},
\begin{displaymath}
\langle(\hat{y}^k-\langle\hat{y}^k\rangle)(\hat{y}^j-\langle\hat{y}^j\rangle)\rangle\,\eta_{jk}<<1,\quad \forall \,y\in\, support(f).
\end{displaymath}
The {invariant warm fluid condition} is manifestly Lorentz invariant and provides a notion of narrowness. In the case of the Minkowski space, it coincides with our narrowness condition.
\end{itemize}

We have emphasized in {\it Section 6} that the approximation $x\to \,\tilde{x}$ is valid not only for short times but also for finite evolution times.  Estimates of when the approximation could not be valid have been given. These estimates are obtained for the worst possible scenario. Thus, it can happens  that the approximation $x(t)\to\tilde{x}(t)$ is valid even for longer evolution times that such estimates indicate.

\subsection*{Acknowledgements} This work was financially supported by EPSRC and Cockcroft Institute at Lancaster University and by FAPESP, process 2010/11934-6 at S\~ao Paulo University and by PNPD-CAPES nº. 2265/2011 at Universidade Federal of S\~ao Carlos.

\footnotesize{
}

\end{document}